\documentstyle[aps,epsfig,floats,prbbib]{revtex} 
\begin{document} 
\begin{twocolumn}
\draft 
\wideabs{
\title{Proximity effects and characteristic lengths
in ferromagnet-superconductor structures} 
\author{Klaus Halterman\cite{klaus} and Oriol T. Valls\cite{oriol}} 
\address{School of Physics and Astronomy and Minnesota Supercomputer 
Institute, University of Minnesota, Minneapolis, Minnesota 55455-0149} 
\date{\today} 
\maketitle 
\begin{abstract} 
We present an extensive theoretical investigation
of the proximity effects that occur in
Ferromagnet/Superconductor ($F/S$) systems. We use a
numerical method to solve self consistently  the
Bogoliubov-de Gennes equations in the continuum.
We obtain  the
pair amplitude and  the local
density of states (DOS), and use
these results to extract the relevant lengths characterizing
the leakage of superconductivity
into the magnet and to study  spin splitting
into the superconductor. These phenomena are investigated
as a function of parameters such as temperature,
magnet polarization, interfacial scattering, sample size
and Fermi wavevector mismatch, all of which turn out to have
important influence on the results. These
comprehensive results should help
characterize and analyze future data
and are shown to be in agreement with  existing experiments.
\end{abstract} 
 
}
\section{Introduction} 

The importance of understanding
the characteristic length scales and
geometrical effects inherent to
heterostructures consisting of
ferromagnets in electrical contact
with  superconductors has received a considerable
reinforcement from the ever increasing
advances in nanofabrication technology.
These advances 
have made it possible (see e.g. Refs.~\onlinecite{kraus,grig,sillanpaa})
to fabricate high quality
nanostructures involving ferromagnets, as well
as normal metals, and superconductors.
In parallel,
there has been significant progress
in the development and refinement
in tunneling spectroscopy
techniques. 
Scanning tunneling microscopy (STM) 
allows one to locally probe the
electronic density of states (DOS) of hybrid systems
over atomic length scales with
sub-meV sensitivity. 

When a 
normal metal is in good electrical contact
with a superconductor,
superconductivity is weakened in the superconductor
and induced in the normal metal.
When the normal metal is
not magnetic, this phenomenon is the traditional\cite{parks} proximity effect
which is described quantitatively via the
pair amplitude, $F(\bf r)$, which encompasses
the spatial dependence of
pair correlations in the both the superconductor
and the normal metal.
If the non-superconductor is a ferromagnet,
the superconducting proximity effect is
drastically modified by the finite exchange field. Furthermore,
the magnetic material can  induce spin polarization in the superconductor,
resulting in a magnetic proximity effect.
The study of the spatial variation
of both the pair amplitude and the local DOS are fundamental to the understanding
of these nanostructures.

When considering such inhomogeneous systems, there are  multiple
length scales involved that must be elucidated.
At $T=0$
the phase coherence in a clean normal metal in contact with a superconductor
decays
inversely with distance from the interface,
with a characteristic length, $\xi_N(0)$,
 that is essentially infinite.\cite{falk}
At finite temperatures,
the phase coherence decays exponentially
over a much reduced distance $\xi_N(T)$.\cite{parks}
Conversely, 
at low temperature,
the pair correlations in the superconductor
become depleted near the interface over
a length scale given by 
the zero temperature superconducting coherence length $\xi_0$,
while
for temperatures
close to $T_c$, Ginzburg Landau 
theory\cite{bdg}
predicts that the depletion is governed by a length
scale $\xi_S(T)$ that 
diverges at $T_c$.
Although the essentials of the standard proximity
effect have been well understood for a long
time,\cite{parks} the length scales
in the intermediate
temperature regimes have
not been systematically or consistently studied
for a  normal metal-superconductor
bilayer system, although
self consistent microscopic calculations exist 
for layered structures,\cite{branko2} and results
within
the quasiclassical\cite{hara} description 
have  been obtained. 
Other predictions
are limited by
being based
on phenomenological or non-self 
consistent approaches.

If the nonmagnetic normal metal is replaced with
a ferromagnet ($F/S$ junctions and structures), the relevant length
scales in the problem are altered significantly. 
Naively, one would expect 
that all phase coherence would be lost in the magnet,
since the superconductor and ferromagnet have opposite
types of long-range ordering: a ferromagnet
favors parallel spin alignment and acts
as an effective pair-breaker, while a superconductor
is  comprised of Cooper pairs with (in the ordinary
$s$-wave pairing considered here) antiparallel spin alignment.
However, a stable superconducting state can arise in the
ferromagnet in which the Cooper pairs have a net center
of mass momentum.\cite{demler}
The spin splitting in the ferromagnet 
introduces a new length scale $\xi_2$ set
by the difference in the spin up and
spin down Fermi wavevectors,
$\xi_2 \propto (k_{F\uparrow}-k_{F\downarrow})^{-1}$,
which is typically much smaller 
than $\xi_0$.

An interesting manifestation of this 
effect is the $\pi$ phase junction comprised
of a ferromagnetic material
sandwiched between superconductors.\cite{blum,prokic,bergeret}
This 
particular
interplay of ferromagnetism and superconductivity
has been studied for some time.\cite{bulaevskii,buzdin82,buzdin92}
The
peculiar oscillatory state
(originally introduced in the context of 
a new superconducting state that arises in
when magnetic
impurities are present\cite{fulde,larkin})
leads, in the sandwich geometry, to a nonmonotonic dependence of the 
critical temperature on the ferromagnet
layer thickness.\cite{radovic,tagirov,fominov2,kontos2}
Other works\cite{ryazanov} have focused on the variation
of the Josephson current with temperature.
For certain values of the exchange field,
spontaneous currents\cite{gyorffy} may 
arise in $F/S$ heterostructures. 

For heterostructure configurations
in which any of the material
thicknesses are of
order of or smaller than the largest of the
relevant intrinsic lengths, size effects
will play a role, and
the finite thicknesses of the 
layers become  important geometric lengths  in the
proximity effects.
It
is then clearly preferable  to tackle the problem
using a theory which does not involve coarse graining over
atomic length scales.
It has been shown that in thin
superconductor-normal metal bilayers,
the interlayer resistance plays a key part.\cite{fominov}
Variations in the local DOS were
calculated as a function of ferromagnet thickness.\cite{baladie,zareyan}
These calculations were all based
on the quasiclassical Usadel\cite{usadel} 
or Eilenberger\cite{eilen} equations.
The use of quasiclassical techniques
may not be appropriate when the
thickness of the materials is only of a few atomic layers.
Also, the Usadel equations are restricted
to the limit when the mean free path is much smaller than
any other relevant length scale in the problem, 
and therefore their use in these situations
is questionable.
It is therefore desirable
to study finite-sized systems using a microscopic,
self-consistent theory that
can accurately account for these geometrical effects.

We are aware of no work that
addresses the 
influence on the proximity effect of
the mismatch between the 
three Fermi energies (or Fermi wave vectors) present
in $F/S$ junctions 
(corresponding to the superconductor, and to the
up and down spin bands in the magnet).
Previous work\cite{tanaka} on this question
was limited to  the case of nonmagnetic metal,
at temperatures near $T_c$. It was found 
that when the Fermi wave vector in the normal
side is smaller than that in the
superconductor,
a strong suppression of the pair amplitude in the
normal metal
ensues.
Also for nonmagnetic materials, the
DOS was studied\cite{branko2}
using a microscopic formalism that
allowed for different Fermi wavevectors,
in the context of 
layered short-coherence length superconducting structures,
but there was no systematic study. 
Therefore this influence
is still an open question in the full parameter range.
For $F/S$ 
junctions, the influence of Fermi wave vector
mismatches 
on the proximity effect
is virtually
uncharted territory. Spectroscopy
studies\cite{zv} revealed a nontrivial dependence of the conductance
spectrum on Fermi wave vector
mismatch, however the proximity effect was ignored there
and the calculation was not self consistent.

Another relevant quantity that
has a strong influence on the
proximity effects and which  has been
insufficiently studied, is the interfacial scattering
strength.
The variation of $T_c$ with interface scattering strength
was calculated,\cite{vodo}
and
the influence of interface scattering was investigated
experimentally\cite{bourgeois,aarts} for 
$S/F/S$
structures.  
High-sensitivity transport measurements\cite{bourgeois}
revealed that
interface barrier strength 
was an important parameter. 
It is thus desirable to study the effects
of varying degrees of barrier strength
on the  characteristic proximity lengths.
Since appreciable scattering at the interface
should lead to a reduction in $F({\bf r})$
near the interface, this is another example
where a systematic, self-consistent solution to
this problem is needed.

The main aim of this paper is therefore to 
present an extensive theoretical investigation
of the influence of these many relevant parameters
on the $F/S$ proximity
effects. We will use for these purposes
a very recently developed numerical 
method that allows for the exact {\it self consistent}
solution of the relevant  microscopic equations.
The approach is based on numerically
solving the continuum Bogoliubov de Gennes (BdG) 
equations\cite{bdg}
for the
quasiparticle amplitudes. The method of numerical self consistent
solution has been described in Ref.~\onlinecite{proximity}
where results  for particular
cases (zero temperature, no barrier or mismatch, and 
semi-infinite geometry)
were given. 
These procedures allow for the study of coherence lengths
much longer than those one can consider in lattice
real-space models.\cite{ting,vecino}.
It was shown\cite{proximity} that 
for $F/S$ junctions there is, besides the usual
characteristic spatial period $\xi_2$,
another length scale $\xi_1\approx\xi_2$ which 
describes  the fast decay 
of $F({\bf r})$ very near the interface.
The above calculations
were performed only at $T=0$. Finding the temperature dependence
of the $F/S$ proximity effects will thus be a part of our task here.

Our objective  is to
investigate the  length
scales characterizing the $F/S$ proximity
effects in 
both bulk and finite sized junctions
consisting of a ferromagnet of varying
polarization,  (including the nonmagnetic
limit) in contact with a superconductor.
As alluded to above, the
often extreme differences in length
scales in the problem require
a self consistent microscopic theory that can deal
with them simultaneously without
the approximations inherent to
quasiclassical and dirty-limit equations.
We shall explore the whole of the parameter range
including the effects
of temperature, Fermi energy mismatch,
interfacial scattering, and finite sample size.
Results will be given for the
pair amplitude and for the local DOS
for both bulk and finite heterostructures, and
thus we will  analyze the various length scales
involved.

Although the objective of this
comprehensive study is to stimulate
new experiments and to help analyze
and characterize the resulting data, we aim also to 
make contact with 
existing experimental work.
Recent STM measurements\cite{courtois} 
indicate a clear modification
to the normal metal density of states for a Nb-Au junction,
as a function of superconductor width.
We compare our results 
with this data using a bilayer model in which
both the normal metal and superconductor
have widths of order $\xi_0$.
We also compare our theoretical results with  tunneling data\cite{sillanpaa}
from
a  $F/S$ 
(Ni-Al) junction
in which
local DOS  
measurements were taken in the superconductor.
The
relatively large exchange energy of Ni makes it an ideal
candidate for investigating
the effect of magnetism on the pairing correlations
in the superconductor. In both cases,
we find, using relevant values of
the parameters describing
the materials used and the geometry
of the experimental set up, very good agreement between
theory and experiment.

The rest of this paper is organized as follows.
In Sec.~\ref{method}, we outline the method of self consistent
solution to the problem. In Sec.~\ref{results}, we present
our results for the
numerous parameters discussed above for different
geometries, and  compare
our results with recent tunnel spectroscopy data. 
Finally in Sec.~\ref{conclusions} we summarize our
conclusions.

\section{Method} 
\label{method}

In this section we briefly outline
our basic equations and methods.
We begin with a brief review of the
spin dependent microscopic
BdG equations in our geometry, and then
outline the numerical method used
for solving them. We  
also explain
the procedure for 
calculating physical quantities paramount in
the study of proximity effects, namely
the pair amplitude and the local DOS. Most of the
techniques used here follow those of Ref. \onlinecite{proximity}.
We will omit many of the details given there 
and we  focus our attention below on those points where 
the methods employed depart from those developed in that work,
such as the inclusion of an insulating barrier and of finite
temperatures.

The BdG equations\cite{bdg} are a conceptually simple and
convenient set of microscopic equations used for studying inhomogeneous
superconducting systems, in our case structures 
involving, in addition to the superconductor,
a ferromagnet or a non-magnetic normal metal. 
We consider a
particular 
slab-like
geometry where the materials are assumed
to extend to infinity in the $x-y$ plane, and
have a total arbitrary thickness $d$ along the $z$ direction,
where the only geometrical 
variation occurs. We denote  the thicknesses
of the ferromagnetic and superconducting layers
by $d'$ and $d-d'$ respectively. 
These materials are in general
separated by a thin insulating barrier at $z=d'$.
Since in this geometry
the system is translationally invariant
in the $x-y$ plane, some aspects of the problem are effectively
one-dimensional. 
For this configuration,
we then have 
two sets of coupled equations, one  for the spin-up 
and spin-down quasiparticle and quasihole wave functions 
$(u_n^\uparrow,v_n^\downarrow)$, and another for
$(u_n^\downarrow,v_n^\uparrow)$. The first 
takes the form\cite{bdg,proximity}
($\hbar=k_B=1$),
\begin{mathletters}
\label{bogo}
\begin{eqnarray}  
\Bigl[-\frac{1}{2m}\frac{\partial^2}{\partial z^2} 
+\varepsilon_{\perp} 
+U(z)-E_F(z)-h_{0}(z)\Bigr] 
u^{\uparrow}_n(z) 
\nonumber \\ 
+
\Delta(z) v^\downarrow_n(z) = \epsilon_n u^{\uparrow}_n(z),
\label{bogo1} 
 \\ 
-\Bigl[ -\frac{1}{2m}\frac{\partial^2}{\partial z^2} 
+\varepsilon_{\perp} +U(z)
-E_F(z)+h_0(z)\Bigr] 
v^\downarrow_n(z) 
\nonumber \\
+ \Delta(z) u^\uparrow_n(z) =  
\epsilon_n v^\downarrow_n(z), \label{bogo2} 
\end{eqnarray} 
\end{mathletters}
where 
$\varepsilon_{\perp}$ is the transverse kinetic 
energy, $\epsilon_n$ are the 
quasiparticle energy eigenvalues
(the index $n$ labels the relevant quantum numbers), 
$h_0(z)=h_0 \Theta(z-d')$ is the magnetic 
exchange energy.
Scattering at the interface (assumed to be spin independent)
is accounted for by
the potential $U(z)=H \delta(z-d')$, where $H$ is the barrier
strength parameter. The pair potential
$\Delta(z)$  satisfies a self-consistency 
condition as discussed below, and
since we will assume that there is no current flowing in the system,
we can take it to be real.
In general, we must allow
for the possibility of having up
to three different Fermi wavevectors or
band widths\cite{zv} in the problem. The quantity $E_F(z)$
equals $E_{FM}$ in the magnetic
side, $0<z<d'$,
so that $E_{F\uparrow}=E_{FM}+h_0$, and $E_{F\downarrow}=E_{FM}-h_0$,
while  in the superconducting side,
$d'<z<d$, $E_F(z)=E_{FS}$.
We will be assuming parabolic bands so that
$\varepsilon_{\perp}=1/2m(k_x^2+k_y^2)$ and there are
three Fermi wave vectors, corresponding to $E_{F\uparrow}$, $E_{F\downarrow}$
and $E_{FS}$.
The solutions for the other set of
wavefunctions  
$(u_{n}^{\downarrow},v_{n}^{\uparrow})$  
are  easily obtained  
from those of Eqns.~(\ref{bogo})
by allowing for both 
positive and negative energies, and then using the
transformation: 
$u_{n}^{\uparrow} \rightarrow v_{n}^{\uparrow},  
v_{n}^{\downarrow} \rightarrow -u_{n}^{\downarrow},  
\epsilon_n \rightarrow -\epsilon_n.$

Equations (\ref{bogo}) 
must be solved in conjunction with the 
self consistency condition for the pair potential, 
\begin{equation}  
\label{del2} 
\Delta(z) =\frac{g(z)}{2} 
{\sum_{n}}'\left[
u_n^\uparrow(z)v^\downarrow_n (z)+
u_n^\downarrow(z)v^\uparrow_n (z)\right]\tanh(\epsilon_n/2T), 
\end{equation} 
where $T$ is the temperature, and
$g(z)$ is the effective BCS coupling constant,
which will be taken to be zero outside the superconductor
and a constant within it.
The prime on the sum in (\ref{del2}) indicates
that the sum is restricted to
eigenstates  
with 
$|\epsilon_n| \leq \omega_D$, where $\omega_D$ is the Debye energy. 

We now solve Eq.~(\ref{bogo})
by expanding the quasiparticle  
amplitudes in terms 
of a complete set of functions $\phi_q(z)$,
\begin{equation}
\label{comset}
u^{\uparrow}_n(z)=\sum^N_{q} u^{\uparrow}_{n q} 
\phi_q(z),  \qquad
v^{\downarrow}_n(z)=\sum^N_{q} v^{\downarrow}_{n q} 
\phi_q(z).
\end{equation}
We will use the normalized particle in a box wavefunctions, 
$\phi_q(z) = \sqrt{{2}/{d}}\sin(k_q z)$, 
as our
choice for the complete set.
Here  $k_q = {q/\pi}{d}$,
and $q$ is a positive integer. 
The finite range of the pairing interaction
$\omega_D$ permits 
the sums in (\ref{comset}) to be
cutoff at an integer $N$ as discussed in Ref.~\onlinecite{proximity},
in a way that depends on the maximum wavevector present.
Upon inserting the expansions (\ref{comset}) into (\ref{bogo}),
and making use of the orthogonality of the $\phi_q(z)$,
we arrive at the following $2N\times2N$ matrix eigensystem, 
\begin{eqnarray} 
\label{nset1} 
\left[  
\begin{array}{cc} 
H^{+} & D \\ 
D & H^{-} 
\end{array} 
\right] 
\Psi_n 
= 
\epsilon_n 
\,\Psi_n, 
\end{eqnarray} 
where $\Psi_n$ is the column
vector corresponding to $\Psi_n^T = 
(u^{\uparrow}_{n1},\ldots,u^{\uparrow}_{nN},v^{\downarrow}_{n1}, 
\ldots,v^{\downarrow}_{nN}).$ 
The matrix elements are given by
\begin{mathletters}
\label{basic} 
\begin{eqnarray} 
H^{+}_{q q'}=\left[\frac{k^2_q}{2m} + \varepsilon_{\perp}\right]\delta_{q q'}+
\int_{0}^{d} dz\, \phi_q(z)U(z)\phi_{q'}(z) \nonumber \\  
 - E_{F \uparrow} \int_0^{d'} dz\, \phi_q(z)\phi_{q'}(z) 
- 
E_{F S}  \int_{d'}^{d} dz\, \phi_q(z)\phi_{q'}(z),\\ 
H^{-}_{q q'}=-\left[\frac{k^2_q}{2m} + \varepsilon_{\perp}\right]  
\delta_{q q'} -\int_{0}^{d} dz\,\phi_q(z)U(z) \phi_{q'}(z) 
\nonumber \\
+E_{F \downarrow} \int_0^{d'} dz \, \phi_q(z)\phi_{q'}(z) 
+
E_{F S}  \int_{d'}^{d} dz \, \phi_q(z)\phi_{q'}(z) ,\\ 
D_{q q'}=\int_{d'}^d dz \,\phi_q(z)\Delta(z)\phi_{q'}(z).
\end{eqnarray} 
\end{mathletters} 
The self-consistency condition is now transformed into, 
\begin{eqnarray} 
\label{selfcon} 
\Delta(z) = \frac{g(z)}{2}
\sum_{p,p'}
{\sum_{n}}' \tanh(\epsilon_n/2T)\times \nonumber \\
\left[
u^\uparrow_{n p}v^\downarrow_{n p'} \phi_p(z)\phi_{p'}(z) 
+
u^\downarrow_{n p}v^\uparrow_{n p'} \phi_p(z)\phi_{p'}(z)\right]
.
\end{eqnarray} 
where the sum over the quantum numbers $n$  encompasses
a sum over the continuous transverse energy $\varepsilon_{\perp}$
and a quantized
longitudinal momentum index $q$,
\begin{equation}
{\sum_n}'\rightarrow{\sum_{\varepsilon_\perp}}'{\sum_q}'.
\end{equation}

The matrix eigensystem Eqn.~(\ref{nset1}) and 
the self-consistency condition (\ref{selfcon})
constitute the primary equations drawn upon in this paper. They are solved
numerically, using the algorithm developed and described
in previous work\cite{proximity}.  The iterative computational 
process is completed
when the maximum relative error in $\Delta(z)$ between successive
iterations is less than a prescribed value as explained below.

Once we have the self-consistently calculated eigenvalues
and eigenvectors,
we can then  construct all relevant physical quantities. For example,
the usual penetration depths are conveniently obtained from 
the pair amplitude $F(z)$,
\begin{equation}
F(z) = \Delta(z)/g(z).
\end{equation}
The pair amplitude, unlike $\Delta(z)$,  is therefore
not restricted by the coupling
constant  to vanish in the non-superconductor.
$F(z)$  gives a quantitative measure 
of the
superconducting correlations in
both the superconductor and non-superconductor where
there may exist 
phase coherence between  particle and
hole wave functions. 
The value of $F(z)$
in the non-superconducting region however,
does not affect the quasiparticle dynamics
since
it is only $\Delta(z)$ that enters into the 
BdG equations.

We can also use our numerical results for the excitation spectra
to calculate the experimentally accessible
local single particle properties
via
the thermally broadened density of
states (DOS) 
\begin{equation}
\label{tdos}
N(z,\varepsilon)= N_\uparrow(z,\varepsilon)+N_\downarrow(z,\varepsilon), 
\end{equation}
where
the local DOS for each  
spin state is given by,\cite{ketterson}
\begin{mathletters}
\begin{eqnarray} 
\label{dos}
{N}_\uparrow(z,\epsilon) 
=-\sum_{p,p'}{\sum_{n}}'
[u^\uparrow_{n p} u^\uparrow_{n p'}
\phi_p(z)\phi_{p'}(z)
 f'(\epsilon-\epsilon_n) 
\nonumber \\
 +v^\uparrow_{n p} v^\uparrow_{n p'}
\phi_p(z)\phi_{p'}(z)
 f'(\epsilon+\epsilon_n)], \\ 
{N}_\downarrow(z,\epsilon) 
=-\sum_{p,p'}{\sum_{n}}'
[u^\downarrow_{n p} u^\downarrow_{n p'}
\phi_p(z)\phi_{p'}(z)
 f'(\epsilon-\epsilon_n) 
\nonumber \\
 +v^\downarrow_{n p} v^\downarrow_{n p'}
\phi_p(z)\phi_{p'}(z)
 f'(\epsilon+\epsilon_n)].
\end{eqnarray} 
\end{mathletters}
Here $f'(\epsilon) = \partial f/\partial \epsilon$ 
is the derivative of the Fermi function.
We will also be  interested in the quantity
\begin{equation}
\label{diff}
\delta N(z,\varepsilon) = N_\uparrow(z,\varepsilon) - N_\downarrow(z,\varepsilon),
\end{equation}
which will be used to characterize the
effective leakage of magnetism into the superconductor.

\section{Results} 
\label{results}
In this Section,
we present numerical results for the
pair amplitude and local DOS,
and discuss other physically meaningful quantities
arising from the self consistent excitation spectra. 
We will  analyze the various length scales
characterizing the influence of the superconductor on the
ferromagnet and vice versa.
Since we will consider
systems with a wide range of superconductor
and ferromagnet widths
and physical parameters, we divide this section into
four different subsections dealing with the following topics:
(1) systematics of the temperature, exchange field, Fermi wave vector
mismatch and  barrier height
for bulk $F/S$ systems.
(2) dependence of the results on the finite thickness
of either the $F$ or the $S$ layer,
and finally (3) a comparison with experimental results.

\begin{table}[t]
\caption{Dimensionless variables}
\label{table1}
\begin{tabular}{ll}
Physical quantity&Dimensionless form\\
\tableline
Exchange energy &$I \equiv h_0/E_{FM}$   \\
Fermi wavevector mismatch & $\Lambda \equiv (k_{FM}/k_{FS})^2$ \\
Temperature&$t \equiv T/T_c$ \\
Coherence length & $\Xi_0 \equiv k_{FS}\xi_0$ \\
Debye energy & $\omega \equiv \omega_D/E_{FS}$ \\
Barrier strength & $H_B \equiv m H/k_{FM}$ \\
Distance relative to interface & $Z \equiv k_{FS}(z-d')$ \\
\end{tabular}
\end{table}

Most of the
results are conveniently expressed in terms of the 
dimensionless quantities
compactly defined and listed in Table \ref{table1}.
Unless otherwise indicated, we
use $\omega$= 0.1 for the Debye cutoff in units of
$E_{FS}$ and
$\Xi_0\equiv k_{FS}\xi_0$=50 in this work. 
All lengths $z$  are measured in 
units of the inverse of $k_{FS}$. For example 
the widths of the ferromagnet or superconductor layers
are represented as, 
$D_F = k_{FS} d'$, and $D_S=k_{FS}(d-d')$. 
The bulk case is studied by choosing
values of $D_S$ and $D_F$ sufficiently large so that
the results become independent of these quantities.

\begin{figure}[t]
{\epsfig{figure=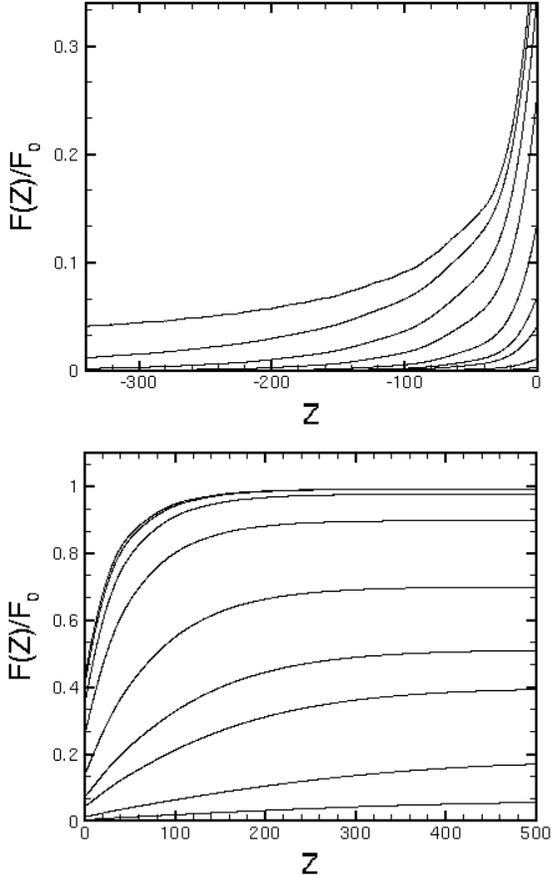,width=.45\textwidth}}
\caption{The pair amplitude $F(z)$,
normalized to the zero $T$ bulk value $F_0=\Delta_0/g$ in the superconductor,
plotted as a function of dimensionless distance
$Z=k_{FS}z$ from the interface. The top panel depicts the normal
metal ($I=0$) region, while the bottom panel shows the superconducting
region. In both cases
the curves correspond, from top
to bottom, to temperatures $t\equiv T/T_c=0,0.2,0.4,0.6,0.8,0.9,0.94,0.98,0.99$.
Note the different vertical
and horizontal scales used in both panels.}  
\label{Fzero}  
\end{figure} 

As mentioned above, we employ a numerical algorithm 
as in Ref.~\onlinecite{proximity} 
to solve the self consistent eigenvalue 
problem Eqns.~(\ref{nset1}), (\ref{selfcon}).
The procedure involves
making a reasonable initial guess for $\Delta(z)$,
where the coordinate $z$ is of course discretized 
for numerical purposes.  The 
initial guess may be taken to be
a previously obtained converged result corresponding
to a slightly different set of parameter
values, or, in  the absence of any such suitable
previous result,  a step function, e.g.,
$\Delta(Z)=\Delta_0 \Theta(Z)$,
where 
Z is the dimensionless
distance from the interface (see Table \ref{table1}), 
and $\Delta_0$ the bulk
value of the gap at $T=0$. We then
diagonalize 
the $2N \times 2N$ matrix
described by Eqn.~(\ref{nset1}) for each value of $\varepsilon_\perp$.
The cutoff number  $N$, as explained in Ref.~\onlinecite{proximity}
depends chiefly on $D \equiv k_{FX} d$, where $k_{FX}$ is the
largest Fermi wavevector in the problem and  $d$ the total thickness,
which is up to $1200 k_{FS}^{-1}$ in our calculations.
We use 5000 different values
of $\varepsilon_\perp$ consistent with the energy cutoff.
The value 5000 is five to ten times
larger than that used
in previous work, which makes for better
convergence and smoother
results. 
Self 
consistency is achieved\cite{proximity} via an iteration process.
The process terminates when the relative error between 
successive 
$\Delta(z)$ is less than a suitable number, chosen
here to be $10^{-4}$ (one tenth of the criterion in previous work).
The pair potential
settles down, after starting
with a step function initial guess,
to its self-consistent form within about twenty five iterations.
This value is typical for most parameter values and system sizes
used in this paper.
The only exceptions are when the temperature
approaches $T_c$ (the bulk  transition temperature
of the superconductor), or when
the superconductor width is
of order of $\xi_0$.
Then
the number of iterations needed for
self-consistency 
is much larger (up to several hundred) if one starts with a step function
guess for $\Delta(Z)$.
This problem can be alleviated  by calculating
$\Delta(z)$ self-consistently for a given temperature
and then use this as input for a nearby temperature
as described above.

\begin{figure}[t]
{\epsfig{figure=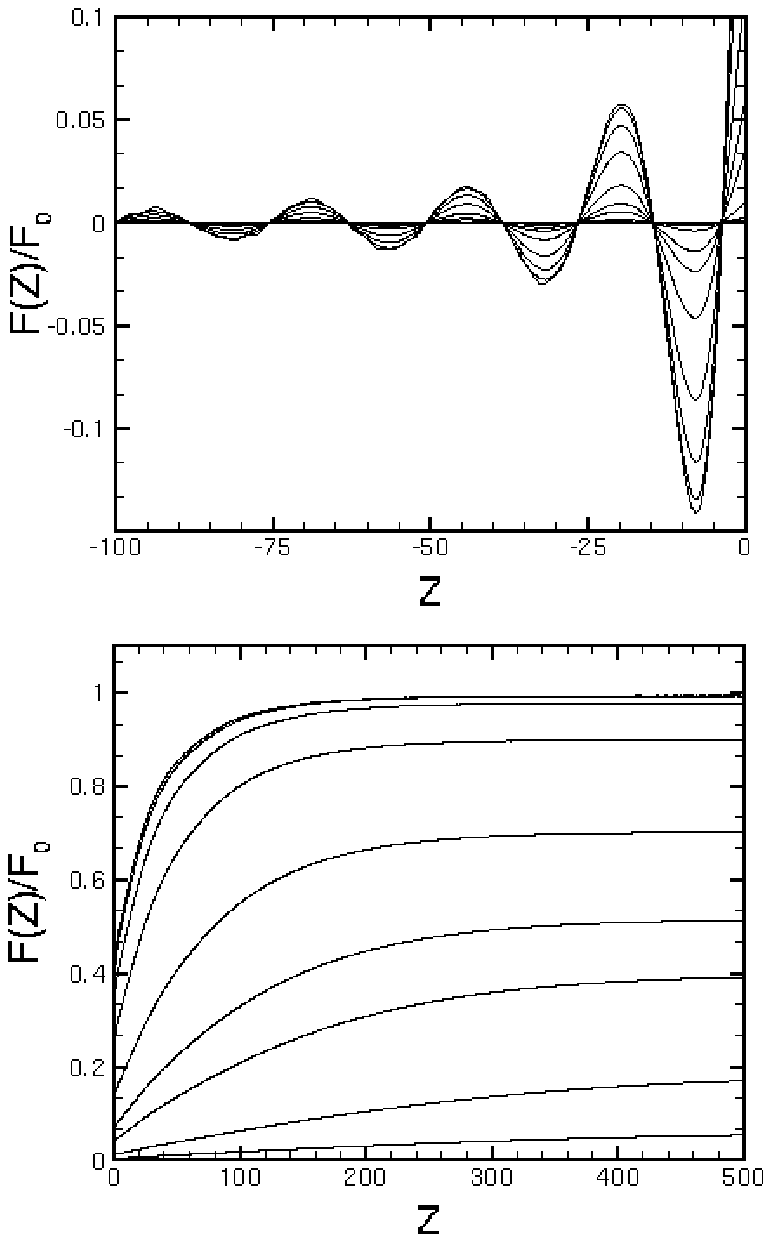,width=.45\textwidth}}
\caption{Normalized pair amplitude as a
function of distance, in both the ferromagnetic (top panel)
and
the superconducting side (bottom panel), at the same temperatures
as in Fig.~\ref{Fzero}. 
The data is for $I=1/4$, with all other parameters
as
in Fig.~\ref{Fzero}. In the top panel, the amplitude of the
oscillations decays monotonically with increasing $t$. }  
\label{F1/4}  
\end{figure} 

\subsection{Systematics of the Parameter Dependence}
\label{bulk}
We consider in this Subsection the dependence of the
results on temperature and on material
parameters (exchange field, wave vector mismatch
and barrier height), in the limit where both ferromagnet
and superconductor are very thick (``bulk'' limit). 
By this we mean that both
$D_F$ and $D_S$ are taken to exceed the 
temperature dependent BCS coherence length.
In this
subsection 
we have taken $D_F = D_S = 12 \Xi_0 $ (recall $\Xi_0=50$),
so that $D_F, D_S \gg \Xi_0$ and we are in the
bulk limit except extremely close to $T_c$, $t\lesssim 0.99$.
We subdivide the analysis
into three categories that address respectively temperature and exchange
effects,
Fermi energy mismatch, and  interfacial 
scattering effects.

\subsubsection{Temperature and exchange energy dependence}
\label{temp}

We now consider how variation of the 
temperature affects the pair amplitude and local DOS
for select values of the dimensionless exchange energy $I$.
To isolate these effects, we assume  that 
there is no interface barrier
and no Fermi energy mismatch ($\Lambda=1$, $H_B=0$, see Table~\ref{table1}).
We first examine the case where the non-superconductor
is a normal metal ($I=0$).
In Fig.~\ref{Fzero}, the pair amplitude $F(Z)$ is shown
in the normal metal and superconductor sides respectively
for a wide range 
of temperatures. In 
all plots, we normalize $F(Z)$ to the zero $T$ bulk value $F_0=\Delta_0/g$.
The two regions $Z>0$ (superconductor)
and $Z<0$ (normal) are plotted in separate panels 
because their significant features 
occur over different
length and vertical scales. The pair amplitude 
however, is continuous across the
interface.
We see that in the normal metal, Fig.~\ref{Fzero} (upper panel), $F(Z)$
has a different functional form 
at zero temperatures (top curve) than at finite temperatures.
At $T=0$  $F(Z)$ has a very
slow decay into the normal region, and
is expected\cite{falk} to decay as the inverse
of the distance from the interface,
\begin{equation}
\label{falkeq}
F(Z) = \frac{c_1}{|Z|+c_2},
\end{equation}
where $c_1$ and $c_2$ are constants.
We find  that the expression Eqn.~(\ref{falkeq}) is valid, but only
in a fairly narrow range close to the interface.
The actual behavior over the range
shown is more complex, which
reflects that the
decay of $F(Z)$  takes place, as we shall see, over two length scales.
Upon increasing $T$, thermal effects reduce
the phase coherence of the electron-hole wavefunctions
and the relevant length scale in the normal metal
is set by\cite{parks}
\begin{equation}
\xi_N(T)=v_{FM}/2\pi T. 
\end{equation}
where $v_{FM}\equiv k_{FM}/m$. It is clear from the
top panel of Fig.~\ref{Fzero}
that as  $T$  increases
the length scale characterizing the decay of $F(Z)$  
decreases. The decay of $F(Z)$ at a fixed, finite temperature
cannot be fit to a single exponential in all of
the spatial region shown.
For temperatures
close to $T_c$, and far from the interface,
an approximate form for the
pair amplitude has been given,\cite{parks}
\begin{equation}
\label{parkstemp}
F(Z) = \Phi(Z)\exp(-|Z|/\xi_N(T)),
\end{equation}
where  $\Phi(Z)$ is a slowly varying function.
Our results agree with Eqn.~(\ref{parkstemp})
in the temperature
regime near $T_c$ ($t>0.9$), and for sufficiently large $|Z|$.
($|Z|>\Xi_0$).
However, 
the overall decay of $F(Z)$ is more complicated
and cannot be
described by a single exponential decay. There always
seems to be a second length scale in the problem, even near $T_c$.

\begin{figure}[t]
{\epsfig{figure=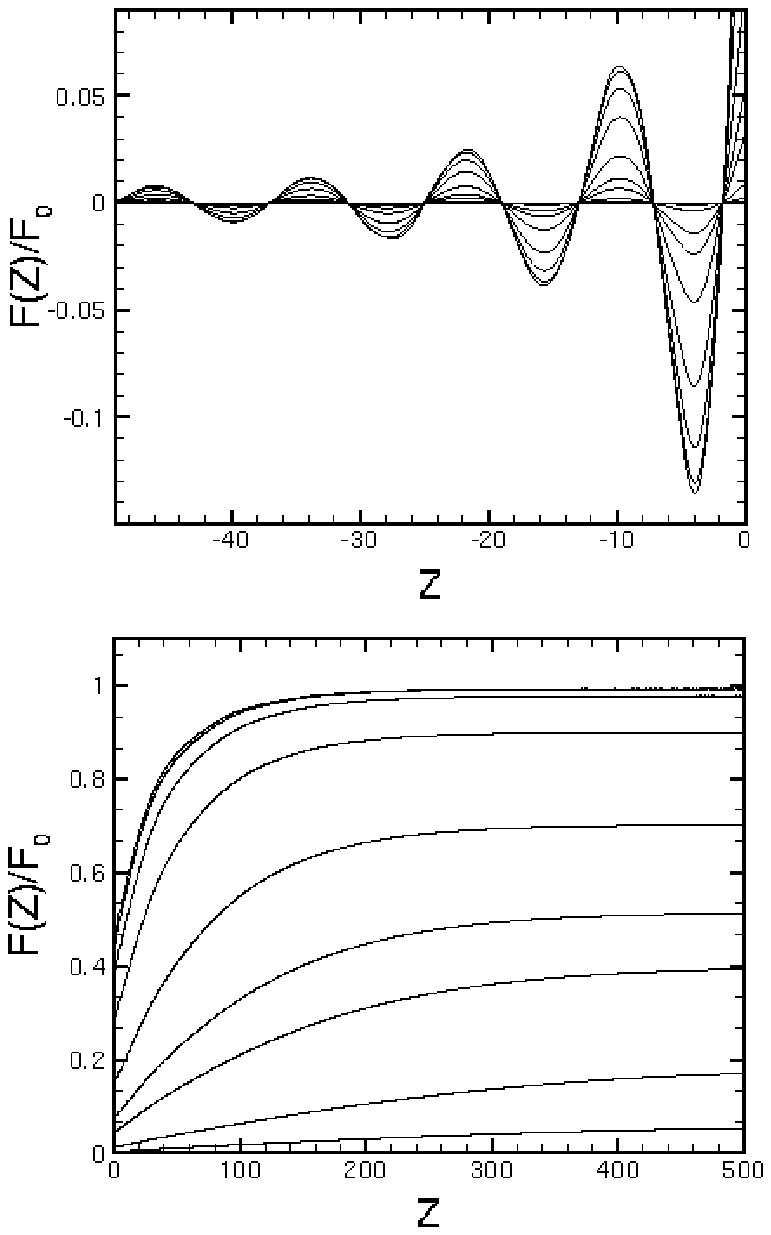,width=.45\textwidth}}
\caption{Normalized pair amplitude for $I=1/2$, plotted
as in Fig.~\ref{F1/4}, for the same parameter values
and temperatures. 
Note the reduction in the characteristic
period compared to Fig.\ref{F1/4}.}  
\label{F1/2}  
\end{figure}

Turning now to the superconductor side, the lower panel
of Fig.~\ref{Fzero} shows
the normalized pair potential
$F(Z)$ at the same values of $T$ used in the panel above it. 
At $T=0$, the characteristic
decay length of the pair amplitude is given by the usual
zero temperature BCS coherence length $\xi_0$. As the temperature
is  increased however, 
the depletion of superconducting correlations
occur over a length scale
which increases with $T$. We denote this scale
(in units of inverse $k_{FS}$) by
$\xi_S(T)$.
For temperatures close to $T_c$ the
profile for the pair amplitude is well known
from standard Ginzburg-Landau theory\cite{bdg},
and has the following form
\begin{equation}
\label{GL}
F(Z)=F_{0}(T)\tanh\left[\frac{Z+Z_0}{\sqrt{2}\,\xi_S(T)}\right],
\end{equation}
where $Z_0$ is a parameter to be determined by the condition
\begin{equation}
\left(\frac{1}{F}\frac{dF}{dz}\right)_{Z=0} = \frac{1}{b},
\end{equation}
and  $b$ is an extrapolation length, which in our
dimensionless units is of order $\Xi_0^2$. These expressions
hold provided\cite{pv}
that $\Xi_0$ is not too small. 
Our results fit Eq.(\ref{GL}) adequately
for temperatures $T\lesssim T_c$
over the entire $Z$ range shown.  
As the temperature is 
decreased, the expression in Eqn.~(\ref{GL})
ceases to be correct for the entire range of $Z$, but
remains an adequate fit within a region of at least one coherence length from
the interface.
We use this expression, therefore
as a fitting
function even at lower temperatures in order
to extract the dimensionless length scale $\xi_S$ characterizing the
decay away from the interface. 
We find that for most of the temperature range,
the characteristic length $\xi_S$  fits well to the Ginzburg-Landau
expression
\begin{equation}
\xi_S(T)=0.74 \,\Xi_0 \left(\frac{1}{1-t}\right)^{1/2}.
\end{equation}

\begin{figure}[t]
{\epsfig{figure=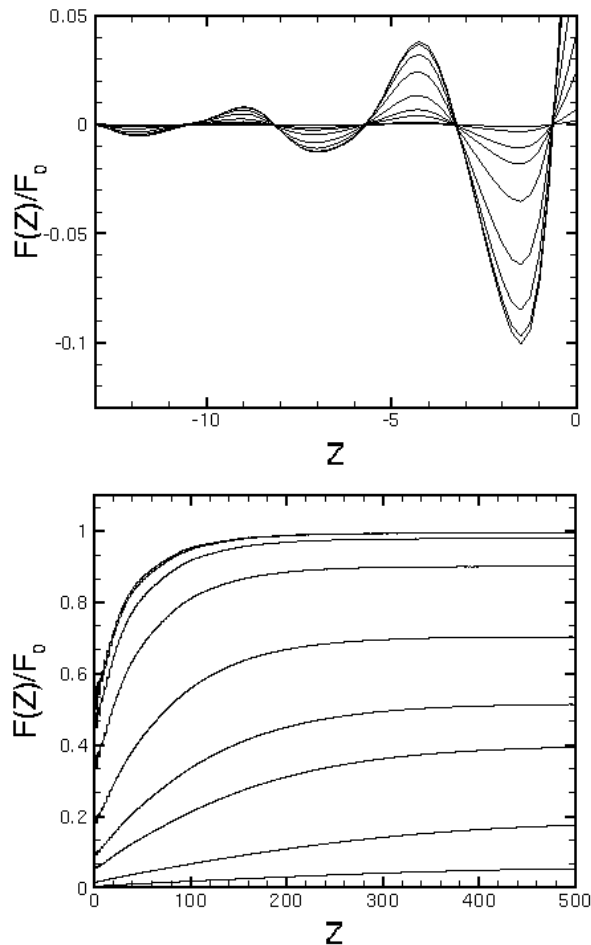,width=.45\textwidth}}
\caption{Normalized pair amplitude for $I=1$, plotted
as in Figs.~\ref{F1/4} and \ref{F1/2}, for the same parameter values
and temperatures. Note the deviation from the simple
sinusoidal decay of Eqn.~\ref{new} on the magnetic side.}  
\label{F1} 
\end{figure}

After 
having shown the dependence of $F(z)$ on
intermediate temperatures $0<T\lesssim T_C$ for $I=0$, and
having verified that our limiting
results are in agreement with previous theory
and expectations for the standard (nonmagnetic) proximity effect, we turn
to the more interesting case where the 
exchange energy parameter $I$ is finite. 
We found above that when $I=0$,
the superconducting correlations extend well
within the normal metal at $T=0$, but decay
more rapidly as the temperature is increased. 
When an exchange field is present, the spin
degeneracy that existed for $I=0$, is removed.
The result is that the Fermi wave vectors of the spin up
and spin down electrons, $k_{F\uparrow},k_{F\downarrow}$, 
are different, and consequently
a Cooper pair entering the ferromagnet
acquires a net center of mass momentum.
The superconducting order induced in the ferromagnet
arises from the
product of particle and
hole wave functions (e.g., $u_n^\uparrow(z) v_n^\downarrow(z)$) 
summed over all
quantum numbers $n$. 
It is the superposition of these wavefunctions that
causes the superconducting
wavefunction to oscillate\cite{demler}
on a length scale set by
the difference
in the spin up and spin down wave vectors,
$\xi_2 \approx(k_{F\uparrow}-k_{F\downarrow})^{-1}$. We have
\begin{equation}
\label{xi2}
k_{FS}\xi_2=\frac{1}{(\Lambda (1+I))^{1/2}}k_{F\uparrow}\xi_2
\end{equation}
where $\Lambda$ is the wavevector mismatch parameter
of Table~\ref{table1}.
Since\cite{proximity} $k_{F\uparrow}\xi_2\approx 1/I$,
we see that the characteristic length of oscillations
scales as $1/I$, and therefore, except at extremely small 
$I$, it is much
smaller than length scale set in the 
normal metal case above.

We have previously studied\cite{proximity}
the explicit form for the pair amplitude
in the ferromagnet
at zero temperature.
We found that for most exchange fields, and except  
extremely near the  the interface,
the pair amplitude  is given by
\begin{equation}
\label{old}
F(Z) = \alpha\frac{\sin[Z/(k_{FS}\xi_2)]}{(Z/(k_{FS}\xi_2))},
\qquad T=0,
\end{equation}
where $\alpha$ is a constant. Very close
to the interface, the pair amplitude monotonically decays
over a characteristic length $\xi_1$ which is
defined as the first point inside the ferromagnet at which $F(Z)$ is
zero. The length scale $\xi_1$ 
goes also as $k_{FS}\xi_1 \approx 1/I$.

We are interested in how the
pair amplitude and
various characteristic lengths
associated with it are modified
as $T$  increases, at finite $I$.
We therefore display in
Fig.~\ref{F1/4} the pair amplitude
at both the ferromagnet and superconductor for $I=1/4$ and
the same  temperature values used in Fig.~\ref{Fzero}.
As in that case, the split panel arrangement is required
by the difference in vertical and horizontal scales, but the function $F(Z)$
is always continuous at the interface. We focus first 
on the ferromagnetic region.
Starting with the top curve in 
the upper panel of
Fig.~\ref{F1/4} ($T=0$),
we see that, beyond a small region of fast decay at
the interface,
the pair amplitude exhibits damped
oscillations, with a
temperature independent period that coincides with the expected
value
$k_{FS}\xi_2 \approx 1/I = 4$, independent of $T$. The envelope decay of
the oscillations varies inversely with distance
as given in Eqn.~(\ref{old}).
The quantity $\xi_1$ is also independent of temperature, since
as can be seen in the Figure, the 
location of the first node of $F(Z)$ as it monotonically goes to
zero near the interface is the same for all temperatures. 
As $T$ increases, however, 
the amplitude of the oscillations in $F(Z)$ markedly decreases.
This decrease, as we shall see below, is not merely a
reflection of the smaller value of $\Delta(T)$ in the bulk superconductor.
Because of this competition between
thermal and exchange energies, the pair amplitude now has a slightly more
complicated functional form than that given by Eqn.~(\ref{old}).
We find that
in order to fit our numerical results, Eqn.~(\ref{old})
must be modified 
by incorporating additional spatial and temperature
dependent factors. The amplitude of the oscillatory decay of $F(Z)$ no longer
decays as the inverse distance from the interface,
but now has an additional slowly varying exponential
term  $\Phi'(Z)$, and
a purely temperature dependent amplitude, $A(t)$,
\begin{equation}
\label{new}
F(Z)=A(t) \Phi'(Z) \frac{\sin\left[(Z+\theta)/k_{FS}\xi_2\right]}{
(Z+\theta)/k_{FS}\xi_2 },
\end{equation}
where $\theta$ is a 
small, weakly $I$ dependent
shift that accounts for the sharp
monotonic decay right at the interface into the
ferromagnet. We find that Eqn.~(\ref{new}) holds for
nearly the entire range values of $I$ $0 \le I \le 1$. Certain
exceptions  occur in the
extreme cases of very small $I\simeq \Delta_0/E_{FM}$
or 
very large $I\simeq 1$ and will be addressed below.
The temperature dependence of the amplitude $A(t)$
in Eqn.~(\ref{new}), 
is fitted well by
the form 
$A(t) = A(0)(1-t^2)$. Thus $A(t)$ decreases faster
with temperature than  the bulk $\Delta(T)$, which
shows that the decrease
of the amplitude with temperature is not merely a normalization effect
but  involves an intrinsic decrease of the pairing at the interface.
Temperature has a marked effect on the amplitude, but
it does {\it not} wash out the oscillations
themselves, which remain quite well defined even
at temperatures quite close to $T_c$.

The superconductor side (bottom panel of Fig.~\ref{F1/4}) shows a
behavior of $F(Z)$ very similar
to that found in the  $I=0$ case, 
with the variation of $F(Z)$ again occurring over
the length scale $\xi_S(T)$.
The effect of the
the exchange field on $F(Z)$ in the superconductor
region 
therefore seems to be minimal at all temperatures. We will see below
however,
that the pair amplitude is only partially useful in
conveying the total effect  of magnetism leakage into
the superconductor. The quantity
$\delta N(z)$ from Eqn.~(\ref{diff}) will be used below
for extracting additional useful information on this question.

Continuing with larger values of $I$, Fig.~\ref{F1/2} shows
the case of $I=1/2$. 
We first
address the ferromagnet side in the top panel.
$F(Z)$  follows the
form given in Eqn.~(\ref{new}), with
a damped oscillatory behavior similar to
that in the $I=1/4$ case. 
However, one should note in making the comparison that
the horizontal axis scale here has been 
reduced by a factor of two with respect to that in Fig.~\ref{F1/4} since
the characteristic
period has approximately halved in accordance with Eqn.~(\ref{xi2}).
In the superconductor side (Fig.~\ref{F1/2}, bottom panel), the 
decay away from the interface is governed by
the length $\xi_S(T)$. The very slight wiggles in $F(Z)$ 
which may be
observed near the interface are due to the increased mismatch
of the two Fermi energy levels in the ferromagnet with $E_{FS}$.
Overall however, the superconducting region 
shows little change compared to the previous case.

\begin{figure}[t]
{\epsfig{figure=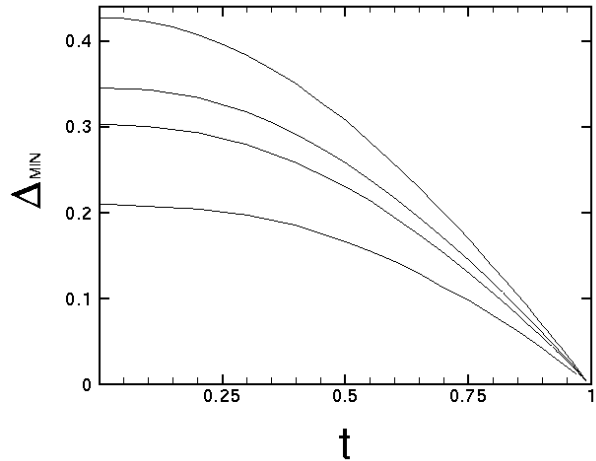,width=.45\textwidth}}
\caption{Variation of the minimum value of the 
self consistent pair potential ($\Delta_{\rm MIN}$)
in the superconductor as a function of dimensionless temperature
for (from top to bottom) $I=0, 0.25, 0.5, 1$.}  
\label{gapvstemp}  
\end{figure}

We now turn to
the extreme (half metallic) case
of $I=1$,
where only one spin band is present in the ferromagnet
at the Fermi level.
We plot results in the same way, and for the same temperatures as
in Figures \ref{Fzero}, \ref{F1/4}, and \ref{F1/2}. 
The top panel in
Fig. \ref{F1}
illustrates the pair amplitude in the ferromagnet.
The characteristic length scale that describes
the main oscillatory behavior is
given from Eqn.~(\ref{xi2}) as  $k_{FS}\xi_2 = 1/\sqrt 2$, 
(recall that we are using $\Lambda=1$ in this subsection). 
The relevant spatial variations occur now only on an atomic scale,
(see horizontal axis). This reflects
that
Andreev processes are inhibited by the absence of Fermi level
down states deep within the magnet.
We also see clear deviations
from the pure damped sinusoidal behavior seen for the
previous two exchange
field values.
The superconducting region (bottom panel)
follows the same pattern as the other cases, but
here we see
that very near the interface
there exist small oscillations of order
of the Fermi wavelength. The oscillations were barely
glimpsed at $I=1/2$ and disappear with 
decreasing $I$. 

\begin{figure}[t]
{\epsfig{figure=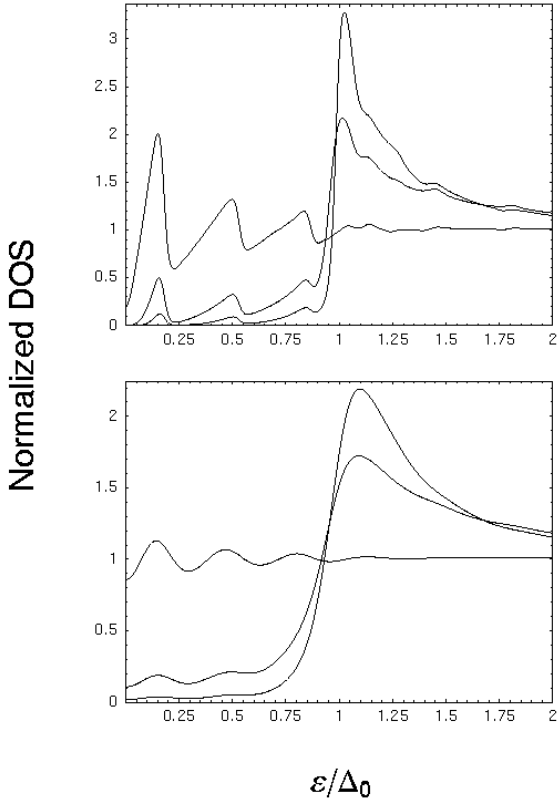,width=.45\textwidth}}
\caption{Local DOS (normalized to its Fermi level value in the normal
state of the superconductor) versus the dimensionless energy
$\epsilon/\Delta_0$ at $I=0$ for $t=0.02$ (top panel) and $t=0.1$
(bottom panel). The curves shown, 
from top to bottom at $\epsilon/\Delta_0<1$, are for
for $Z=-100,100,200$ respectively.}
\label{dos6}
\end{figure}

The pair amplitude at the interface
at constant temperature  decreases markedly with $I$,
while at constant $I$ it decreases with $T$. 
We illustrate this in Fig.~\ref{gapvstemp},
where $\Delta_{\rm MIN}$ is the minimum
value of the normalized $\Delta(Z)$ in the superconductor. 
This minimum occurs right at the interface, and
because of 
the relatively wide horizontal scale 
in the bottom panels of Figs.~\ref{Fzero},\ref{F1/4}, and \ref{F1/2},
it is not really possible to read its value
from these Figures. 
We see in this Figure
that the effect of the exchange
field is quite pronounced, and that  as the temperature
approaches $T_c$, all curves tend to collapse into a nearly straight line
tending to zero.
The depletion of superconducting correlations 
with exchange field
at 
the interface is also quite evident from the data shown.

Having studied the spatial dependence of
the superconducting correlations, it is
now pertinent to examine the local density of states
at various positions on both
sides of the interface.
The local DOS gives  further direct insight into
the proximity effect, and more important, it is an
experimentally accessible quantity. Its calculation 
is achieved through Eqn.~(\ref{tdos})
and the computed self consistent spectra. 

We again consider first 
the case  $I=0$. The top panel in Fig.~\ref{dos6} shows the normalized
(all results for the DOS in this work
are presented normalized to the  superconductor's normal state DOS
at the Fermi level)
local DOS
at $t=0.02$, while the panel below it
corresponds to  $t=0.1$.
For both cases the numerical results are plotted for the local DOS at
the positions $Z=-2\Xi_0,2\Xi_0,4\Xi_0$  that is, one position in the
normal metal and two in the superconductor.
Focusing  on the very low
temperature case, (top panel), we consider first the normal 
metal region at $Z=-100$ (top curve in the subgap region). 
We see there the sawtooth-like pattern
characteristic of the de Gennes-St.~James states
as predicted long ago\cite{stjames}. The DOS 
is small but finite  at the Fermi
energy due to  filling by thermally excited
quasiparticles, and
then rises
nearly linearly at small energies. 
The Andreev bound states are illustrated by the 
peaks in the DOS. These are due to constructive
interference of the electron and hole wavefunctions,
as they undergo Andreev reflection 
at the F/S interface and  normal reflection
at the vacuum-normal metal interface at the opposite end of the sample.
The characteristic
energy $E_c$ of the peaks is determined by 
adding up the phases for a given trajectory\cite{bruder}.
It can be seen that, in agreement with theoretical expectations, the 
first peak occurs at an energy 
$E_c \approx \pi v_{FM}/4 d' \approx 0.2 \Delta_0$,
while the other peaks occur 
approximately at multiples of $2 E_c$.
The energy scale $E_c$ can  be seen directly in the  calculated 
self-consistent spectrum.\cite{thesis}
In the energy region
below the gap ($\epsilon/\Delta_0 <1$), we find that
for nearly longitudinal momenta 
($\varepsilon_\perp \approx 0$), there exists roughly
three excitation branches, at the same energies as
the peaks seen in the upper panel of Fig.~\ref{dos6}.
These peaks subsequently broaden
due to the numerous quasiparticle states
with momenta nearly perpendicular to the interface.
The two other curves in the top panel of Fig.~\ref{dos6}  
correspond to points in the superconductor
at $Z=2 \,\Xi_0$ and $Z=4 \,\Xi_0$. The
bound states clearly flatten out as one moves further
into the superconductor,
and there are no longer any states at
the Fermi energy.
The BCS gap becomes quite evident
at the position $Z=4\,\Xi_0$, when only a hint of nonvanishing
DOS can be seen at  energies below $\Delta_0$. 

\begin{figure}[t]
{\epsfig{figure=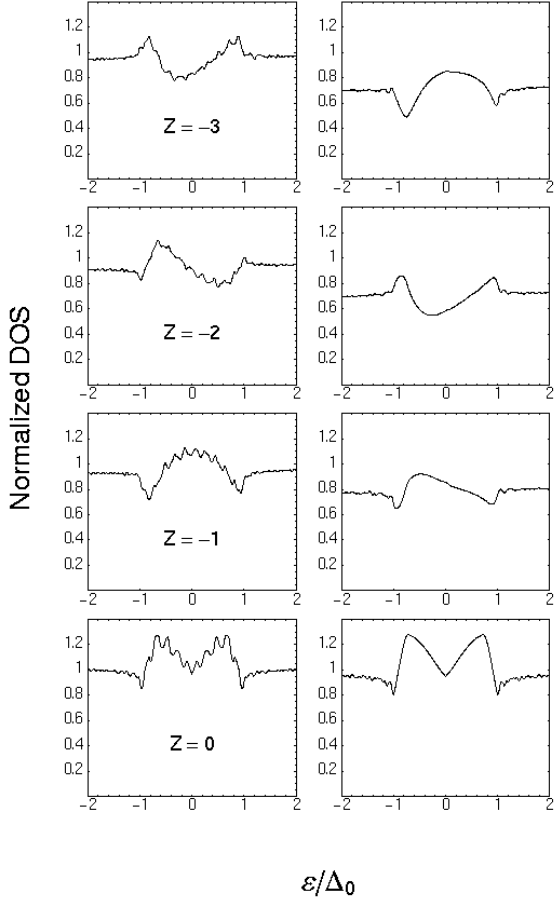,width=.45\textwidth}}
\caption{Local  DOS  (normalized as in Fig.~\ref{dos6})
in the ferromagnet for $I=1/2$ and $t=0.02$ (left column)
at four positions near the interface.
The right column corresponds to $I=1$ and the
same temperature and positions used
in the left column.}  
\label{dos1/21}   
\end{figure}

Increasing the
temperature tends to smear the previous lower
$T$ results. Figure~\ref{dos6} (bottom panel) 
illustrates this for the case of $t=0.1$, and
the same positions as the panel above it.
In the normal metal,
the sharp pronounced zig-zap pattern is now smoothed
into a series of  humps. The same is true
for the DOS in the 
superconductor, where
thermal excitations
fill regions inside the gap, while flattening and spreading
out the smoothed
BCS peaks. 

We now can proceed to study the effects of a finite exchange field
$I$ on the local DOS in the superconductor and ferromagnet.
We begin with the ferromagnet region at $I=1/2$, and
$t=0.02$.
As already seen in Ref.~\onlinecite{proximity},
the effect of $I$ on the local DOS is very drastic.
The bound state phenomena in the normal side
relatively far from the interface are no longer
observable at $I=1/2$ 
because the overall decay of superconducting correlations
takes place over considerably smaller distances.
On the other hand, this decay
takes place now in a nonmonotonic matter, which gives
rise to a new set of features in the ferromagnet, very
near the interface:
in the  left column of
Fig.~\ref{dos1/21} we show the 
(normalized) DOS at four positions
at and very near  the interface.
The influence of the oscillatory
pair amplitude (Fig.~\ref{F1/2}) becomes evident as we examine
the four plots in the this column.
The 
subgap structure in the top curve ($Z=-3$) 
evolves so that maxima and minima become
reversed at $Z=-1$, closer to the interface. 
Comparing with Fig.~\ref{F1/2},
we see that the oscillating superconducting order 
has in effect induced oscillations
in the local DOS as a function of position within the ferromagnet,
and that the large exchange field induces
noticeable particle-hole asymmetry.
The length scale at which the DOS flips coincides with the
characteristic
distance $k_{FS}\xi_2$, given
in Eqn.~(\ref{xi2}).
The first and third panels in Fig.~\ref{dos1/21} (left)
are separated by $\Delta Z = 2$. 
Although this Figure depicts results obtained for  a rather
low temperature, the oscillatory behavior is never
completely washed out by the temperature, as remarked above
in conjunction with the discussion of $F(Z)$.

\begin{figure}[t]
{\epsfig{figure=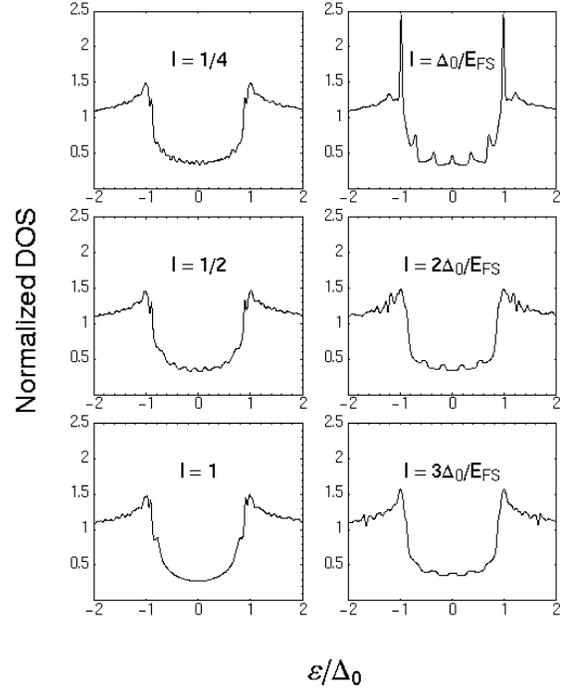,width=.45\textwidth}}
\caption{Normalized local DOS at $Z=\Xi_0$ in the superconductor
and $t=0.02$
for various
exchange fields as labeled in each panel. The left column corresponds
to intermediate and large exchange fields, and the right 
column to small values of $I$ as indicated.}  
\label{dosall}   
\end{figure}

The same behavior holds true {\it a fortiori} in 
the right column of Fig.~\ref{dos1/21}, 
which displays data 
for the half metallic case $I=1$, at the same 
temperature and locations. One important difference
between this and the left column is the spatial scale
at which the DOS oscillations occur. Since $1/I =1$, the 
complete DOS inversion should occur at points separated
by an interval $\delta Z$ of order
unity. Indeed, the curves in the right column
of Fig.~\ref{dos1/21} reflect 
this.

In Fig.~\ref{dosall} we consider the superconductor
side of the junction. In the left column we illustrate the local DOS 
one coherence length $\xi_0$ from the interface for three
different exchange fields ranging from $I=1/4$ to $I=1$,
at temperature t=0.02. 
In
the top panel  ($I=1/4$)
there is a wide
U-shape opening for energies $|\epsilon/\Delta_0|<1$.
The  opening then starts 
to get smaller and
the curve trends upwards
for $I=1/2$, as
seen
in in the middle plot of this left column. There is also
a slight decrease in the number of states
at the Fermi level ($\epsilon=0$). These
results are consistent with those
previously obtained\cite{proximity} at
zero temperature. For the half-metallic case $I=1$,
the left bottom panel 
shows a further slight reduction in states
at $\epsilon/\Delta_0=0$, but but there is barely a hint
of the asymmetry found in the ferromagnetic side.

It is of interest also to study the 
case where the exchange field parameter $I$ is weak, 
of the order  of $\Delta_0/E_{FS}$. 
Possible resonance effects have been predicted
to occur at these small exchange energies.\cite{fazio}
Results are shown in the right
column of Fig.~\ref{dosall}. For the values of the parameters
considered in this subsection, it follows
from Table~\ref{table1} and the BCS
relation $k_F\xi_0=(2 E_F)/(\pi \Delta_0)$ 
that $\Delta_0/E_{FS}=0.0127$ here.
We consider the same temperature and location in the superconductor
as for the larger values of $I$ in the left column of the Figure.
We start with the case $I=\Delta_0/E_{FM}$,
(top curve in this 
column). Focusing on energies
$|\epsilon/\Delta_0|<1$, we see 
a dramatic sharp 
peak in the DOS near the gap edge, and five
smaller peaks at lower energies. Upon doubling $I$, 
(second panel from top) the sharp peak structure
near $\epsilon/\Delta_0=1$  vanishes, and there
are now four subgap small peaks. Indeed, we have
found that the sharp peak exists only at $I=\Delta_0/E_{FM}$.
Finally, the bottom curve shows
that for $I=3\Delta_0/E_{FM}$, 
only three small subgap peaks remain.
It would
be of considerable interest to verify
experimentally the appearance and disappearance of the very sharp peak
at the gap edge for $I=
\Delta_0/E_{FM}$.

\begin{figure}[t]
{\epsfig{figure=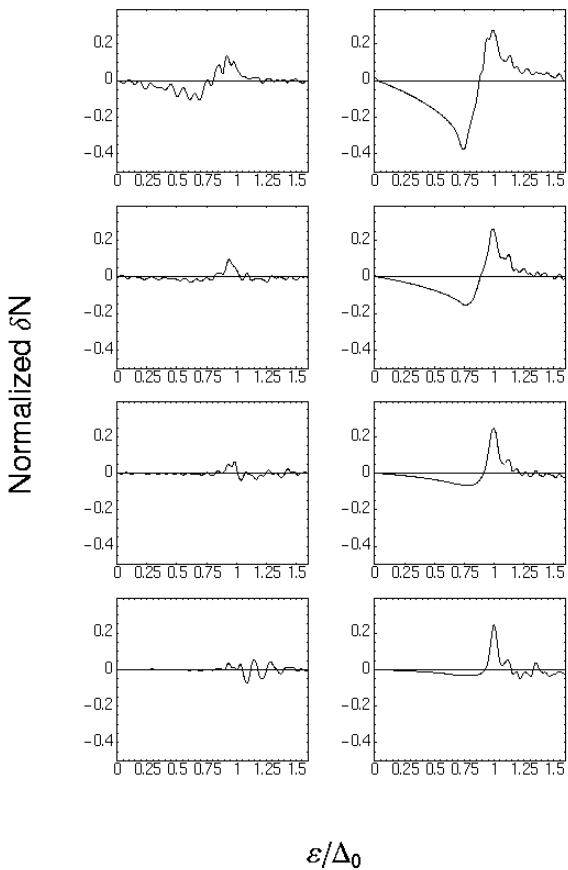,width=.45\textwidth}}
\caption{$\delta N$ (normalized to the normal state
DOS summed over spins)
in the superconductor for I=0.25 (left column) and I=1 (right column) 
at $t=0.02$ and positions (from top to bottom) $Z=n\Xi_0, n=1,4$.}  
\label{diffdos2}   
\end{figure}

Examination of the bottom (superconductor side) panels
of Figs.~\ref{Fzero},~\ref{F1/4},~\ref{F1/2}
and~\ref{F1}, shows that the  exchange field 
seems to affect the pair amplitude $F(Z)$ within the superconductor 
relatively little over any significant length scales.
We want to study the possibility, however, that
the differential local DOS
$\delta N(\epsilon,z)$, defined in Eqn.~(\ref{diff})
may show, within the superconductor,
magnetic penetration  over distances much
larger than that revealed by $\Delta(z)$.
We have a hint that this might be the case:
the results for $\Delta(z)$ in Ref.~\onlinecite{proximity},
exhibited no significant dependence on $I$, while
$\delta N(\epsilon,z)$
was appreciably nonzero within a small region in the
superconductor near the interface. These previous results were obtained
for the special case where $k_{FS}=k_{F\uparrow}$, a 
condition, which corresponds to an $I$ dependent mismatch parameter
$\Lambda=1/(1+I)$, that may yield results different from
the case $\Lambda=1$ considered here.

We  examine in
Fig.~\ref{diffdos2},
the normalized $\delta N(Z)$ for a field parameter values $I=1/4$ (left
column) and $I=1$ (right column). 
We use $t=0.02$ 
and choose four locations  in the superconductor, at $Z=n\Xi_0, n=1,4$.
At the position $Z=\Xi_0$, 
(top panels) there is a clear manifestation of the magnetic
proximity  effect through a
nonzero value of $\delta N$ near
$\epsilon/\Delta_0=1$. The effect decreases
as $Z$ increases and, for $I=1/4$ it nearly dies out 
at $Z=4\,\Xi_0$, that is, after several coherence lengths.
At $I=1$ (half metallic case) the effect is more prominent
and extends over larger distances. 
However, the integral of $\delta N$  over
energies turns out to 
be always extremely small (as we shall see  
below) at these distances, at which only the self consistent energy
spectral distribution shows magnetic penetration spin-splitting effects.
We see that $\delta N$ vanishes at the Fermi level
but the details of this fairly long range redistribution of energy states 
are nontrivial and difficult to interpret.
Nevertheless, that the effect is larger near the gap energy can be
readily understood if one recalls that\cite{btk} the imaginary
part of the wave vector of injected quasiparticles below the gap
(in a non-self consistent approach) vanishes as the gap edge is approached.

In previous work\cite{proximity} it was found
as mentioned above, that for  $I=1$
and no mismatch between the spin up and superconductor band 
($E_{F\uparrow}/E_{FS}=1$),
the effect of the exchange field on the superconductor was 
small and $\delta N(Z)$  decayed away over a few atomic distances.
However the current
condition $\Lambda=1$ implies, at $I=1/4$ and particularly at $I=1$
a considerable mismatch between $k_{FS}$ and $k_{F\uparrow}$.
We shall study this point in detail below, in the context of
our discussion of wave vector mismatch in general. 

\begin{figure}[t]
{\epsfig{figure=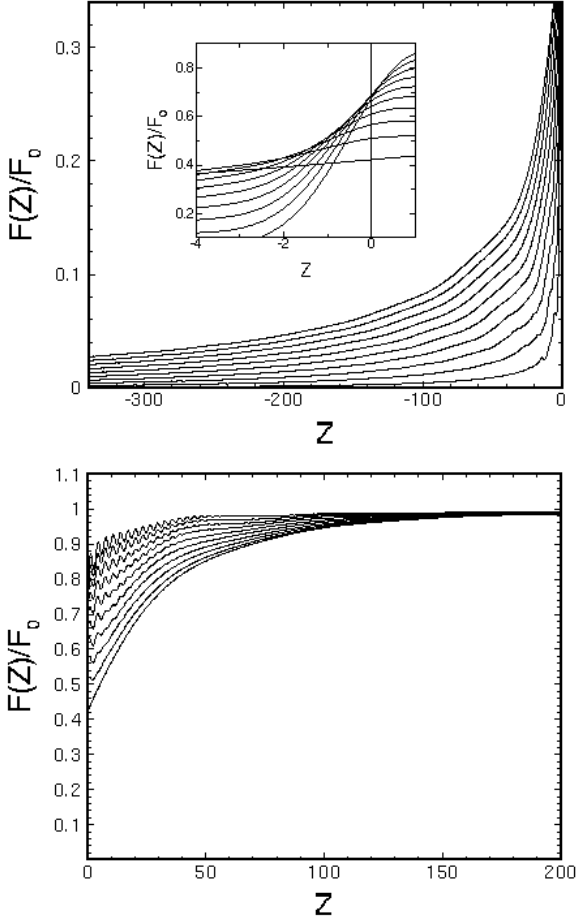,width=.45\textwidth}}
\caption{$F(Z)$ at $t=0.1$
and $I=0$, plotted vs. $Z$,
for  values of the mismatch parameter $\Lambda$ (see Table~\ref{table1})
$\Lambda=0.1-1$,
in increments of $1/10$. 
Top panel is the normal metal region: the curves,
from top to bottom, correspond to decreasing $\Lambda$. The 
bottom panel is for the superconductor, and
curves from top to bottom correspond to
increasing $\Lambda$.  Inset:
crossing of the curves close to the interface (vertical line) on the normal
side
emphasizing that $F(Z)$ is continuous.}  
\label{F6fwm}   
\end{figure}

\subsubsection{Fermi wavevector mismatch}

In the last paragraph, we have seen  that,
(as
previously\cite{zv} seen in a different context)
mismatch among the three Fermi wave vectors
involved in the problem (or the three band
widths) may have a considerable effect on the results.
As such mismatch is experimentally unavoidable, 
we now proceed to investigate it in some detail.
Thus, we will consider values of the 
mismatch parameter $\Lambda$
(Table \ref{table1}) different from unity. We will still
keep the interface barrier parameter at $H_B=0$.

In Fig.~\ref{F6fwm} we show the pair amplitude
for $I=0$, $t=0.1$, and  $\Lambda$ varying from
the previous case of unity down to 0.1 in increments of $1/10$.
We focus on the situation
where the bandwidth in the normal metal
is smaller than that of the superconductor. 
This is the more common situation in  $F/S$ structures and in
any case, it turns out to lead to more prominent effects.
The top panel in Fig.~\ref{F6fwm} shows 
an overall suppression of superconducting
correlations with decreasing $\Lambda$ (increasing mismatch).
Extremely near the interface,  $|Z|\approx 1$,
(see inset),
$F(Z)$ drops rapidly and the curves cross 
as  $\Lambda$ increases.
At $\Lambda=0.1$ (bottom curve in this main panel) the phase coherence
is virtually destroyed at the
the distance of
$Z=-200$.
Away from the interface, the pair correlations still
decay  in accordance with 
Eqn.~(\ref{falkeq}), the only modification being
a mismatch dependent amplitude factor $g(\Lambda)$,
\begin{equation}
F(Z)=\frac{g(\Lambda)}{|Z|+c_2},
\end{equation}
where $g(\Lambda)$ is an increasing function of  $\Lambda$.
Thus,
a smaller bandwidth in
the normal metal tends to restrict the
influx of Cooper pairs. Physically, since the parallel momentum 
of a Cooper pair at
interface is conserved, the longitudinal component 
is restricted
by the smaller number of states accessible in the normal side.\cite{tanaka}
This is consistent with the bottom panel of Fig.~\ref{F6fwm}
which shows $F(Z)$ in the superconductor.
The top 
curve, corresponding
now to $\Lambda=0.1$, shows a smaller characteristic
length of decay from the interface 
than that for the $\Lambda=1$ case (bottom curve).
The subsequent value of $\Delta(Z)$ at the interface, $\Delta_{\rm MIN}$, 
decreases smoothly as $\Lambda$ increases. The effects of values of 
$\Lambda$ in the range $\Lambda>1$, (not shown) are in the opposite direction,
but always much less prominent. For this reason this range has
been deemphasized.

The general trends are  similar in the ferromagnetic case.
The top panel of Fig.~\ref{fwm5} displays 
the damped oscillations
of $F(Z)$ in the magnet, at $I=1/2$. 
The period of the damped oscillations varies inversely with $\sqrt{\Lambda}$, 
in agreement with  Eqn.~(\ref{xi2}), but  they nearly wash out 
when $\Lambda=0.1$.
Also, the sharp monotonic decline very near the interface
increases in slope with greater $\Lambda$, so that
$F(Z)$ first reaches zero at a
greater distance from $Z=0$, thus also increasing  $k_{FS}\xi_1$
defined earlier.
Quantitatively, one can obtain an excellent 
fit for the damped sinusoidal dependence of the pair amplitude 
by using 
Eqn.~(\ref{new}) with  $k_{FS}\xi_2$ as a fitting parameter.
The results of doing this yield values
in excellent agreement with Eqn.~(\ref{xi2}).

In the bottom panel of Fig.~\ref{fwm5},
the 
top five curves show the
drop in $F(Z)$ within the superconductor as the interface
is approached. 
The main feature that stands out is 
that the results in the range $0.7<\Lambda<1$ are nearly
independent of $\Lambda$, while those for $\Lambda<0.7$ exhibit
a marked $\Lambda$ dependence, similar to that seen
in the bottom panel of Fig.~\ref{F6fwm}. This unexpected result arises 
as at $I=1/2$ 
and $\Lambda=2/3$ 
one reaches the special
point where $E_{F\uparrow}=E_{FS}$. This property is further exemplified
in the inset where we present $\Delta_{\rm MIN}$
as a function of $\Lambda$. One can see a kink
in the curve at about $\Lambda=2/3$. 

\begin{figure}[t]
{\epsfig{figure=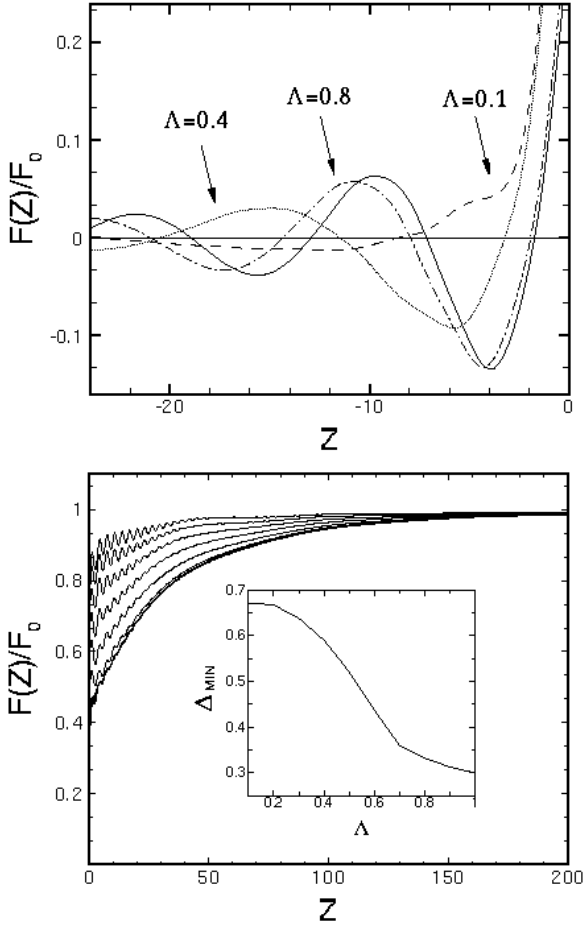,width=.45\textwidth}}
\caption{Normalized pair amplitude 
for the case $I=1/2$ and  $t=0.1$.
The top panel illustrates the
variation of $F(Z)$ for three different
$\Lambda$ (as indicated) in the ferromagnet, while in the 
superconductor side
(bottom panel) results for the same
$\Lambda$ values used in Fig.\ref{F6fwm} are shown (in order
of decreasing $\Lambda$ from top to bottom).
The inset depicts the value of $\Delta(Z)$ at the interface,
$\Delta_{\rm MIN}$ as a function of the parameter $\Lambda$}  
\label{fwm5}  
\end{figure}

The Fermi wavevector mismatch influences also
the local DOS as shown in the following figures.  
Starting again with $I=0$, we present in Fig.~\ref{dos6fwm}
the normalized DOS in (left column) the normal metal at $Z=-100$ for 
three different
values of  $\Lambda \neq 1$, and (right column) for the
superconductor at $Z=1\,\Xi_0$, at the same $\Lambda$ values.
The corresponding $\Lambda=1$ results are in Fig.~\ref{dos6}.
Both cases are at the low temperature of $t=0.02$.
As $\Lambda$ decreases, 
(higher mismatch) we see that, for $\epsilon/\Delta_0<1$, the
bound state peaks decrease, until they disappear at $\Lambda=0.1$.  
The decrease at higher energies reflects our normalization.
The superconductor side (right column)
shows an interesting  trend. We examine there the point
$Z=50$, one coherence length away from the interface. As one decreases
$\Lambda$, the small but distinct peaks within
the gap turn into small wiggles at $\Lambda=0.4$ and disappear
altogether for $\Lambda=0.1$. The BCS peaks
at the gap edge become much more pronounced, indicating
a substantial reduction in Andreev reflection 
at the interface because of the increased
mismatch in Fermi energies, which results in
superconductivity being  more ``confined" to the superconductor.

We saw in Fig.~\ref{fwm5} that 
the damped oscillations of the pair amplitude
inside the magnet increase in wavelength and decay quicker with 
decreasing $\Lambda$ (increasing mismatch).
We investigate the effect this has on the DOS in Fig.~\ref{dos5fwm}
for the same value of $I$, $I=1/2$, as in that Figure.
The left column of Fig.~\ref{dos5fwm} depicts 
the changes in the magnet side 
local DOS associated with the same variation in $\Lambda$ presented
for  $I=0$ in Fig.~\ref{dos6fwm}.
Corresponding results at $\Lambda=1$ were given in Fig.~\ref{dos1/21}.
As expected the DOS  again experiences oscillations  correlated with 
the characteristic length $k_{FS}\xi_2$, 
as a function of  $\Lambda$. 
The left arrangement of panels in
Fig.~\ref{dos5fwm} illustrates this
point. The coordinate is fixed to $Z=-4$.
One can see an evident inversion between the 
$\Lambda=0.8$ and $\Lambda=1$ (Fig.~\ref{dos5fwm}) cases, whereby
the positions of minima and maxima are interchanged.
The superconductor side is examined in the right column,
which shows the same
$\Lambda$ values as in the left set of panels, at a distance $Z$ 
of one correlation
length inside the material. 
The top  panel
exhibits  behavior similar to that found for $\Lambda=1$:
the density of states within
the gap is appreciably nonvanishing, and the 
peaks at $\epsilon/\Delta_0=1$ are
relatively low. The peaks near the gap edge for $\Lambda=0.4$,
below are more prominent
and there is a concomitant decrease in subgap states. This is
quantitatively different from what we saw at  $I=0$, where
there were more subgap states and the BCS peaks
were significantly sharper. Finally, the bottom panel
reveals a near absence of states below the gap, and 
the usual BCS-like peaks at $\epsilon/\Delta_0=1$.

\begin{figure}[t]
{\epsfig{figure=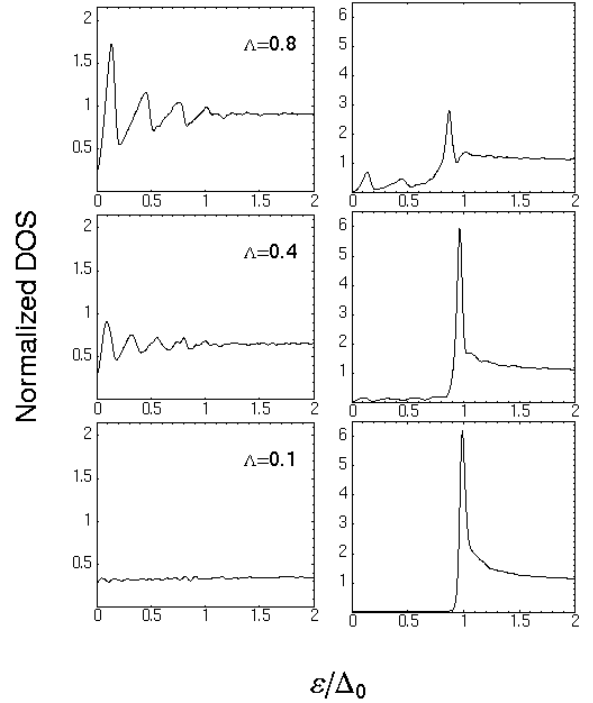,width=.45\textwidth}}
\caption{Normalized local DOS 
plotted versus the dimensionless energy $\epsilon/\Delta_0$ at
$I=0$, and $t=0.02$ for three values of the mismatch parameter 
$\Lambda$ labeled in the left panels. The position is
at $Z=-2~\Xi_0$ in the
normal metal side (left column), and at $Z=\Xi_0$
in the superconductor side (right column).}  
\label{dos6fwm}  
\end{figure}

\begin{figure}[t]
{\epsfig{figure=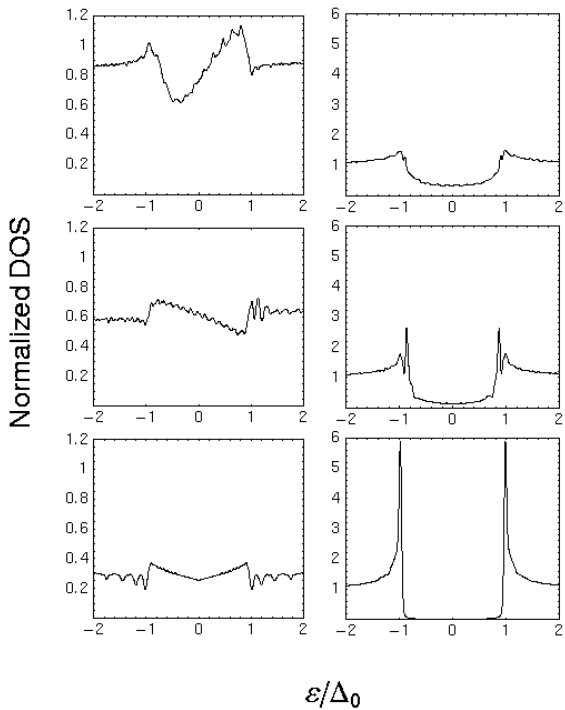,width=.45\textwidth}}
\caption{Normalized local DOS for $I=1/2$, and
with the same temperature and $\Lambda$ values used
in Fig.\ref{dos6fwm}. The left column
corresponds to the position $Z=-4$ in the ferromagnet,
while the right column corresponds to $Z=\Xi_0$ in the
superconductor.}  
\label{dos5fwm}    
\end{figure}

It was also seen in Fig.~\ref{fwm5} that $F(Z)$ in the superconductor
decayed away from its bulk value near the interface
in a  strongly  $\Lambda$-dependent manner.
To address whether 
this parameter also affects the
spin-splitting 
in the
superconductor, we now calculate $\delta N(z,\epsilon)$
defined  in Eqn.~(\ref{diff}).
Figure~\ref{diffdosfwm} shows $\delta N$ (still at $t=0.02$
and normalized as in Fig.~\ref{diffdos2})
as a function
of the dimensionless energy
and at a distance
of one $\xi_0$ from the interface, for several
values of  $\Lambda$, at $I=1/2$.
Starting at $\Lambda=1$, we see 
an effect reminiscent of what was seen in Fig.~\ref{diffdos2}: there is
a net negative spin population for $\epsilon/\Delta_0\lesssim 0.85$, 
then for larger energies, a greater
number of up spin states, which decays quickly so that
the two spin states equalize for $\epsilon/\Delta_0>1.5$.
Next, consider the case where 
$\Lambda=2/3$  (when $k_{F\uparrow}=k_{FS}$).
In agreement with Ref.~\onlinecite{proximity} the result
is nearly zero for this special value.
This value of $\Lambda$ is also the  point
at which the $F(Z)$ plots (Fig.~\ref{fwm5}) start diverging
with further 
decreases in $\Lambda$.
At this special matching point little leakage of magnetism into
the superconductor occurs.
The importance of this crossover point becomes more evident
in the remaining curves, where
the mismatch parameter is decreased to $\Lambda=0.4$
and then to 0.1. 
The sign of the $\delta N$ variations
with energy is reversed.
This pattern, and the relatively large maximum 
and minimum values of $\delta N$  
reflect that the high peaks reached by the DOS at these values of
$\Lambda$ (see Fig.~\ref{dos5fwm}, right column), occur at slightly
different values for the  up and down spin bands.
Again, the magnetic moment at those distances is very small:
if one integrates the normalized $\delta N$ over the variable
$\epsilon/\Delta_0$ the result is of order $10^{-2}$
at $Z=\Xi_0$, changing sign at $\Lambda=2/3$. Only
very near the interface, at values of $Z$ of order unity,
we find that this integral is larger, and of course
always positive.
As a rule, spin-splitting effects in the self consistent DOS 
extend through several times $\xi_0$ (being larger near $\epsilon/\Delta_0=1$
for the reasons already discussed), except of course at the reversal point.
The parameter characterizing the degree of wavevector mismatch
is therefore important in the study of proximity effects
on both sides of the $F/S$ interface.

\begin{figure}[t]
{\epsfig{figure=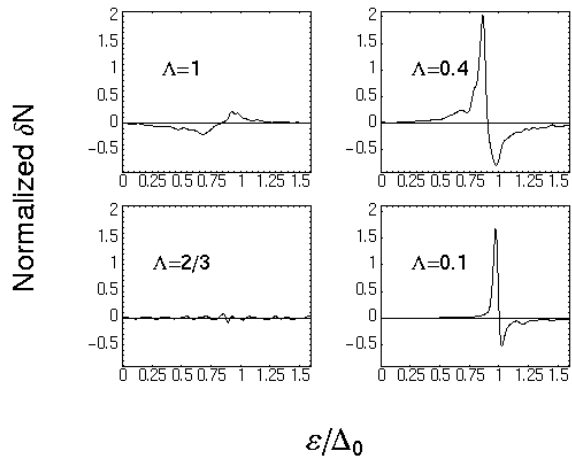,width=.45\textwidth}}
\caption {Normalized $\delta N$ [see Eq.\ref{diff}],
at $Z=\Xi_0$ in the
superconductor for different values of $\Lambda$.}  
\label{diffdosfwm} 
\end{figure}

\subsubsection{Interface scattering}

Up to this point we have considered only transparent interfaces.
A thin oxide layer at the interface adjoining a
superconductor and a normal metal or ferromagnet can be
modeled by  a repulsive delta function
potential as defined earlier in Sec. \ref{method}. The spin
independent scattering
strength is parameterized in  dimensionless units 
by the quantity $H_B$, defined
in Table.\ref{table1}.

\begin{figure}[t]
{\epsfig{figure=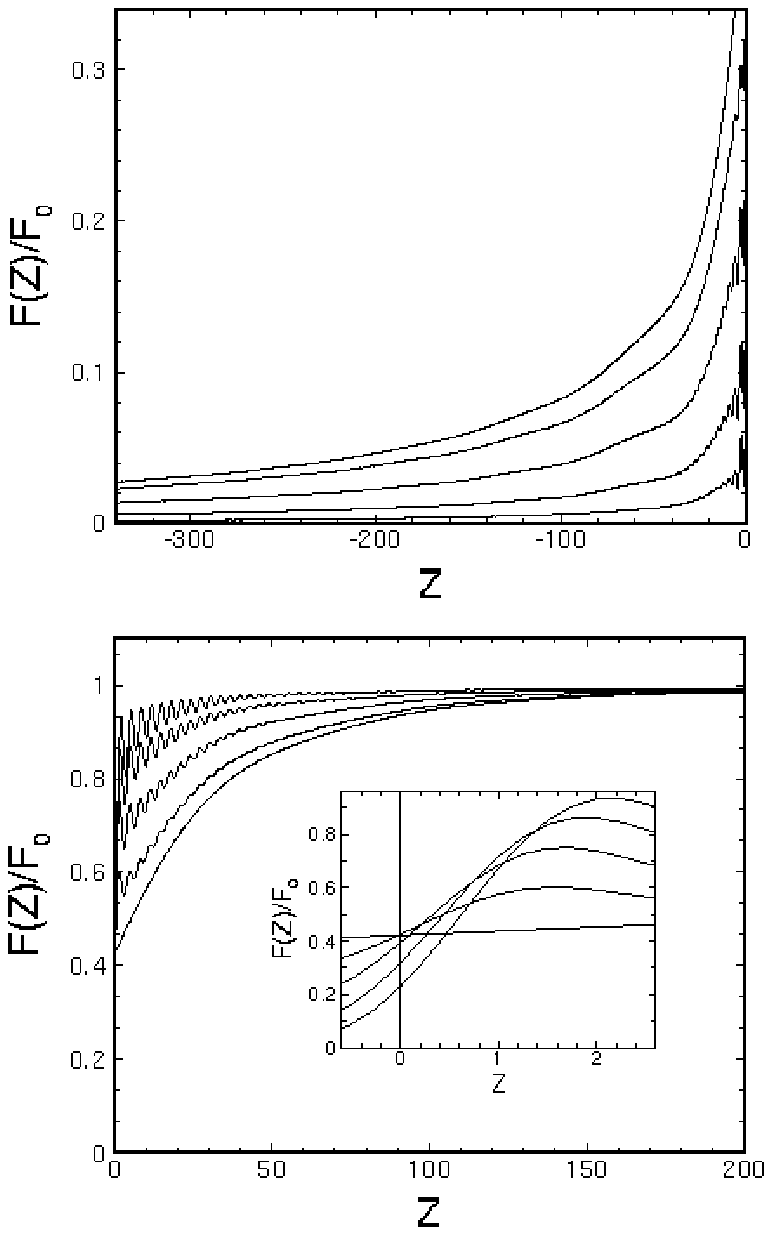,width=.45\textwidth}}
\caption{The normalized pair amplitude at $I=0$ 
for different
values of the barrier strength (see Table~\ref{table1}),
$H_B=0,0.2,0.4,0.6,0.8$,
and $t=0.02$. The top panel is for the normal metal
region. The curves, from top to bottom,
correspond to increasing $H_B$. The bottom panel depicts the superconductor 
side. The curves from top to bottom are in order of decreasing
$H_B$ in this region.
The inset reveals the crossing of the curves near the interface (vertical line) 
at length scales too small to be seen in the main panel.}  
\label{F6z}  
\end{figure}

We fix the temperature to $t=0.1$ and the mismatch parameter
to unity for this study.
In Fig.~\ref{F6z} we present the pair amplitude
in both the superconductor
and the normal ($I=0$) metal, in the format of previous Figures. 
We consider
five equally spaced values of the dimensionless 
barrier strength ranging from 
$H_B=0$ to a relatively strong 
interfacial scattering
barrier $H_B=0.8$.
First we examine the normal metal side, in the top panel.
The effect of the barrier is quite pronounced,
as the pair amplitude still decays
slowly into the normal side, but with an overall large decrease
in amplitude. Also evident are Friedel type\cite{sv} oscillations
in $F(Z)$ near the interface:
as the insulating barrier becomes stronger, the two parts
of the system become more isolated from each other.
The pair amplitude in the region shown is  adequately
fit by the functional form of Eqn.~\ref{falkeq} but
with the parameters $c_1$ and $c_2$  being 
both functions
of $H_B$.
Within the superconductor, the pair correlations
within a range of order $\xi_0$ from
the surface  increase with increasing
barrier strength. This is illustrated in 
the bottom panel of Fig.~\ref{F6z},
where the rise of $F(Z)$ near the interface can be
seen to sharpen with increasing $H_B$.
The top curve ($H_B=0.8$) has the least overall
variation in the scale shown. The 
oscillations near the interface
have the same period as in the normal metal, and their
amplitude increases with  $H_B$.
As remarked in connection with Fig.~\ref{F6fwm},
$F(Z)$ is continuous at the
interface, but here however, the curves
cross 
in the superconductor very near the interface.
This is illustrated in the inset since
this property is not
visible in the horizontal scale of the main figure, where
we emphasize longer-range changes.

\begin{figure}[t]
{\epsfig{figure=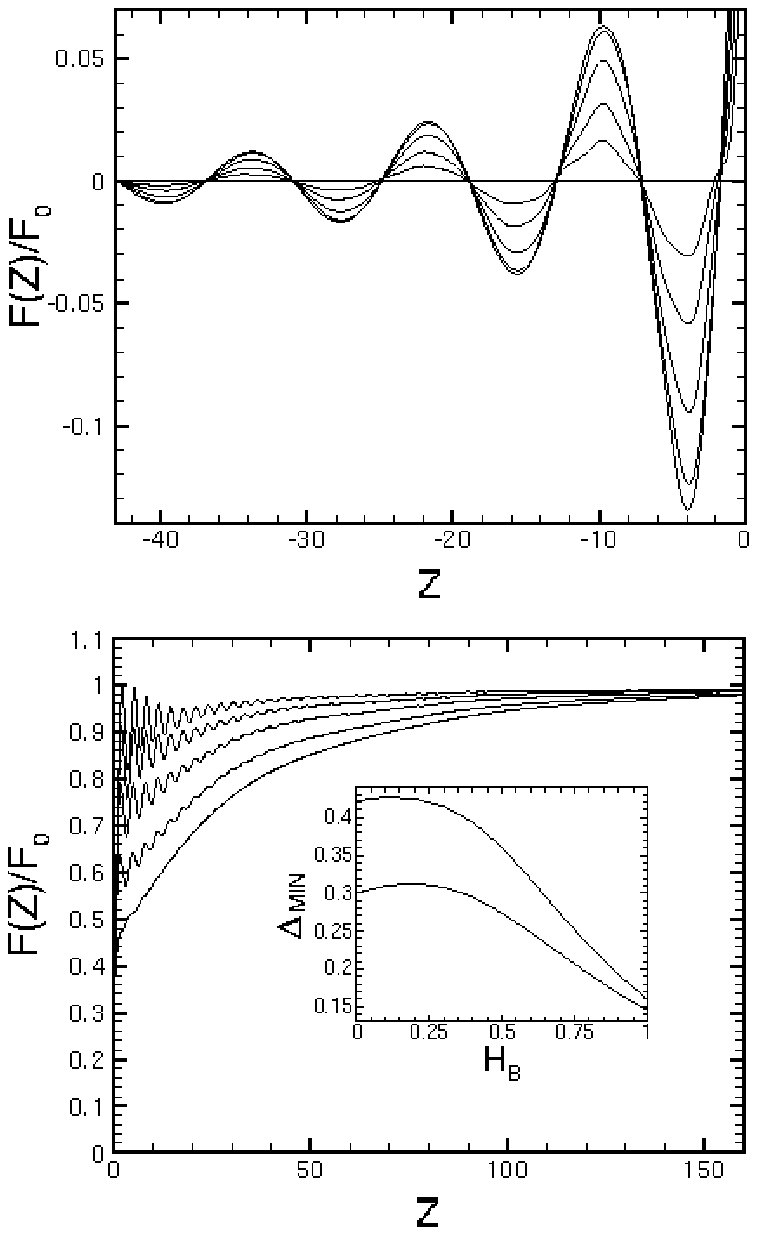,width=.45\textwidth}}
\caption{Normalized pair amplitude at $I=1/2$,
for the ferromagnet (top panel), and the superconductor (bottom panel). 
All other parameters
and curve trends are the same as in Fig.\ref{F6z}. The inset illustrates
the variation of $\Delta_{\rm MIN}$ as a function of the
barrier strength $H_B$.}  
\label{F5z}    
\end{figure}

The case of a finite exchange field (with $I=1/2$)
is shown in Fig.~\ref{F5z}.
All other parameter values are the same as in Fig.~\ref{F6z}.
Examining first the magnet side,  (top panel)
reveals that  the amplitude of the damped 
oscillations decreases
as the scattering potential
is increased. 
The period is independent of $H_B$, in agreement with Eqn.~(\ref{xi2}). 
The additional
decay of the amplitude of the oscillations can be incorporated
into Eqn.~(\ref{new}) through a multiplicative factor that
decreases linearly with $H_B$.
The location of the first node of $F(Z)$ in the magnet
is  nearly unaffected, demonstrating that both the
characteristic length scales $k_{FS}\xi_1$ and $k_{FS}\xi_2$
are independent of interface transparency. 
On the superconductor side, the bottom panel
shows how the length scale over which
$F(Z)$ regains its bulk value from the interface decreases
as the scattering potential increases.
We again see  oscillations in $F(Z)$ for finite values
of $H_B$, near the interface, here  they
are more marked than in the $I=0$ case. The 
value of $\Delta_{\rm MIN}$ at a given $H_B$ always 
decreases with $I$. 
This  is illustrated in the inset at
the bottom of Fig.~\ref{F5z},
where we plot $\Delta_{\rm MIN}$ as a function
of $H_B$, for two values of $I$. The difference between the
two curves is largest at $H_B=0$, while 
they trend closer with increasing barrier strength. 
As the barrier becomes very strong  the proximity effects 
become minimal.

\begin{figure}[t]
{\epsfig{figure=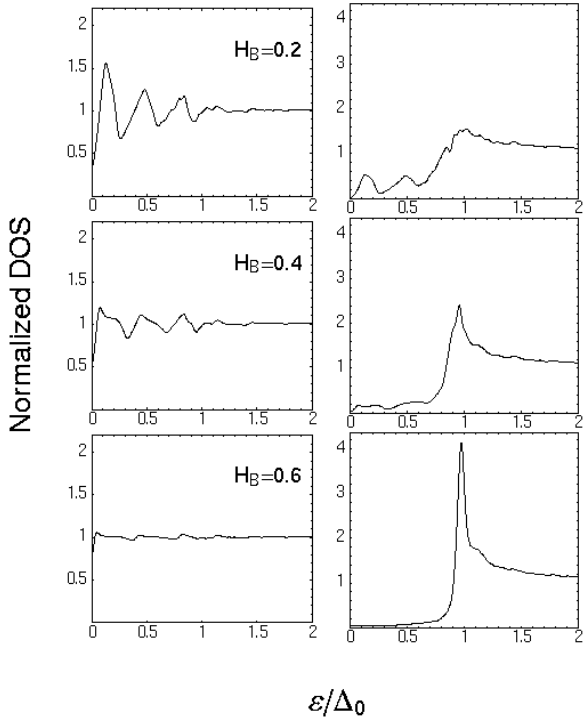,width=.45\textwidth}}
\caption{The effect of a finite barrier on the local DOS:
Normalized local DOS at $I=0$ for finite values
of the barrier strength $H_B=0.2,0.4,0.6$
and temperature $t=0.02$. The left column corresponds to
$Z=-100$ in the normal metal, and the right column is 
for $Z=\Xi_0$. The value of $H_B$ at each panel on the right
is the same as that in the panel on its left.}  
\label{dos6z}  
\end{figure}

We conclude this subsection with a look at how the local
DOS is modified due by finite barrier strength. 
In Fig.~\ref{dos6z}, we
display the  normalized 
local DOS for  $I=0$, at $t=0.02$.
The left column 
shows the results for the normal metal side.
The curves all correspond to $Z=-100$. 
The $H_B=0$ results are in Fig.~\ref{dos6},
while here we show results for increasing values of $H_B$
as labeled in the figure. The general trend on increasing
the scattering potential is a reduction in the magnitude of the peaks
for subgap energies. The characteristic energy spacing $E_c$ shows
relatively little change, but
the shape of the peaks is drastically altered. 
The corresponding change in the DOS profile 
for the superconductor
is  shown in
the right column, where we present the local DOS for
the same values of the barrier strength and at the point $Z=\Xi_0$. 
The de Gennes St. James bound states
still evident in the top curve become smeared out 
until at the bottom curve, where $H_B=0.6$,
the influence of the normal metal becomes
almost nonexistent. Thus we find that although both
the insulating  barrier and the Fermi energy mismatch  tend to
destroy superconducting order in the non-superconductor,
their DOS signature is quantitatively different.

\begin{figure}[t]
{\epsfig{figure=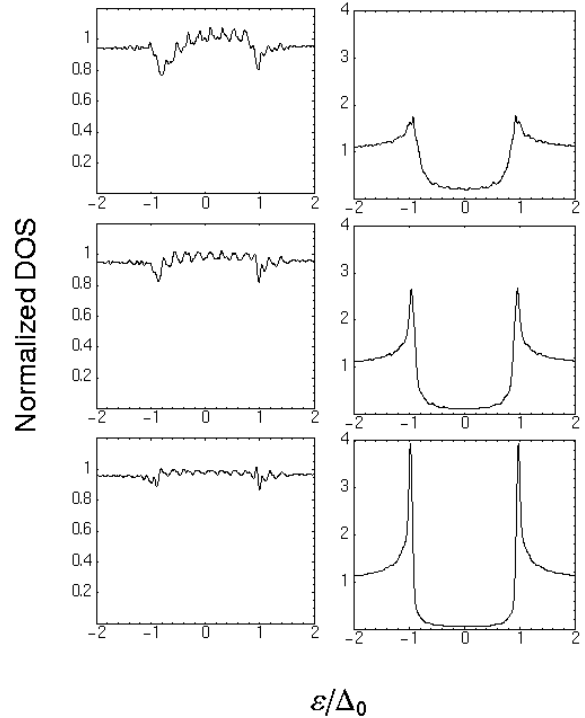,width=.45\textwidth}}
\caption{Local normalized DOS for $I=1/2$ and the same barrier strengths
and parameter values used in Fig.\ref{dos6z}. The
distance on the left side, however, is $Z=-4$ (left column) while
on the superconducting side (right column) we have still $Z=50$.}  
\label{dos5z}  
\end{figure}

The effect of  barrier
strength at finite exchange fields 
can be seen
in Fig.~\ref{dos5z}. We 
take $I=1/2$, with all other parameter values at the same values as 
in Fig.~\ref{dos6z}. 
In the ferromagnet side, the top left curve
in Fig.~\ref{dos5z}
demonstrates a wide
structure in the subgap DOS. Upon increasing the barrier
strength (lower panels), there is a dramatic reduction in this
structure. 
As can be seen, when the barrier strength is rather large ($H_B=0.6$),
the DOS shows very minimal signs of the
proximity effect at the distance from the interface considered.
The superconductor side (right column), for $Z=\Xi_0$ and
the same values of $H_B$ as the other panel, reflects the
trend seen in the ferromagnet. The subgap
states at zero barrier (top curve, see also Fig.~\ref{dosall}) 
gradually disappear
with increasing $H_B$, while simultaneously 
sharp BCS peaks develop. For $H_B\ge0.5$, the
results follow closely the $I=0$ case (see Fig.~\ref{dos6z}(b))
since the influence of the non-superconductor material has vanished.
These results illustrate the importance of fabricating samples
with good, clean interfaces.

\subsection{Structures}
\label{structures}

\begin{figure}[t]
{\epsfig{figure=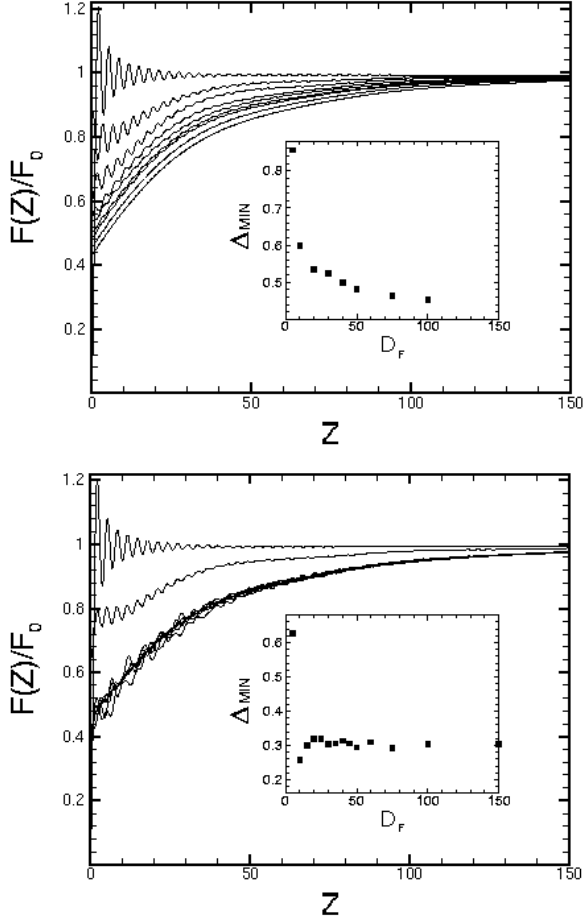,width=.45\textwidth}}
\caption{Pair amplitude in a superconductor in proximity
to a non-superconducting layer of finite thickness $D_F$ (see text). The top
panel shows results for $I=0$. The curves from top to bottom correspond
to $D_F=0,5,10,20,30,40,50,100,200$, respectively. The bottom panel has
results for $I=1/2$. 
The top
two curves are for $D_F=0,5$. The other curves, all of which
essentially coincide, are for the remaining values of $D_F$ as shown
in the top panel. The insets show $\Delta_{\rm MIN}$ vs $D_F$
in each case.}  
\label{nvary}  
\end{figure}

All of the above results pertained to ``bulk'' structures, in that
both slabs were taken
to have dimensions significantly larger than the zero temperature
BCS coherence length.
We now address what happens when either the
ferromagnet or the normal metal
is thin enough so that size effects are appreciable. 
A bilayer system of this type
is an appropriate model for the case when the mean free path in
the finite layer is
larger than the layer's width.
We will present a broad range of results,
varying $D_F$ from a few atomic spacings
up to of order $\Xi_0$, while keeping $D_S\gg\Xi_0$,
and vice versa. 
We will consider the case where both $D_F$ and $D_S$
are small in Sec.\ref{compare}.
For the sake of brevity, we will take the interface to be
transparent ($H_B=0$), the mismatch parameter
to be $\Lambda=1$, and fix the temperature to $t=0.02$.

\begin{figure}[t]
{\epsfig{figure=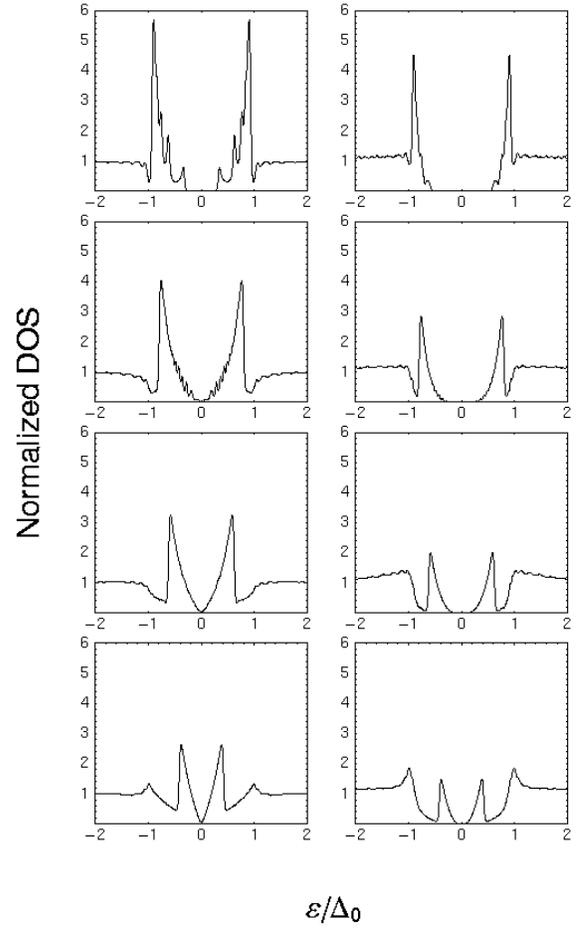,width=.45\textwidth}}
\caption{Local DOS for the same geometry analyzed
in Fig.~\ref{nvary}, at $I=0$. From
top to bottom, each set of panels corresponds
to $D_F=20,50,100,200$.
The left column shows the normal metal ($I=0$)
local DOS $N(z,\epsilon)$, spatially
averaged over one $\Xi_0$
as described in the text. The right column is the local DOS at $Z=50$
in the superconductor.}  
\label{dosnvary}  
\end{figure}

We begin with a normal metal ($I=0$) of finite width
backed by a ``bulk'' superconductor,
taken here to be $D_S = 16 \Xi_0$.
The top panel in
Fig.~\ref{nvary}
shows the pair amplitude in the superconductor for
various normal metal widths. We show $F(Z)$ only 
for the superconductor
side since the pair amplitude in the finite
normal side is cut off at different distances.
The top curve corresponds to a single superconductor slab (zero
width for the normal metal),
while subsequent curves are for increasing normal metal widths
ranging up to $D_F=200$.
The oscillations  near the interface at zero
or very small $D_F$, are again the well-known geometrical
Friedel oscillations.
The largest changes in the pair amplitude
near the interface occur for $0<D_F<20$. 
When $D_F\ge50$, the
characteristic length scale for superconducting
depletion is given approximately by
the coherence length $\xi_0$. 
The inset displays 
$\Delta_{\rm MIN}$ (as usual, the value 
of $\Delta(Z=0)$) as a function of $D_F$. It is seen
that $\Delta_{\rm MIN}$ drops rapidly until about
$D_F=50=\Xi_0$ and thereafter it decays more slowly.

The same geometry as in 
the top panel of Fig.~\ref{nvary}
but with the finite non-superconducting layer
being ferromagnetic ($I=1/2$) is considered in the
bottom panel of this Figure, which again depicts the
pair amplitude in the superconductor.
As the ferromagnet thickness $D_F$ begins to increase from zero
the pair amplitude drops very rapidly, as  in the
$I=0$ limit. A notable distinction exists here however. 
When $D_F$ is larger than about $D_F=10$, the 
characteristic length scale over which $F(Z)$
rises to its bulk value becomes approximately independent of $D_F$.
This behavior is also seen in more detail in the inset of the same Figure
where we plot $\Delta_{\rm MIN}$ as a function of $D_F$.
The decay of $\Delta_{\rm MIN}$ occurs
nearly entirely in the region  $D_F\le 10$, while for $I=0$ it takes
place over a much more extended range. This of course reflects that the
superconducting penetration (at low $T$) into the normal metal is very large,
while for a magnet with $I=1/2$ it is characterized by a length of order
$\xi_2$.
Once $D_F$ reaches that limit, further increases are ineffective. 

We have also calculated the DOS for the 
geometries used in Fig.~\ref{nvary}.
Since the non-superconductor 
layer is in some cases quite
thin ($D_F<\Xi_0$), the local DOS in the
normal region
exhibits strong oscillations
as a function of  $Z$.
For this reason, we present results
for the spatial average of $N(Z,\epsilon)$
over a distance in the $Z$ direction equal to 
the layer thickness $D_F$ if $D_F<\Xi_0$, or over one 
dimensionless coherence length, $\Xi_0$,
if $D_F>\Xi_0$.
In the latter case
this average is centered at $Z=-D_F+\Xi_0/2$. We 
present in the left column of Fig.~\ref{dosnvary}, the
averaged DOS within the normal ($I=0$) metal for
four different thicknesses,
at  $t=0.02$. 
The top left panel corresponds to
a thin film with
$D_F=20$ (recall the superconductor
is  in the bulk limit). A clear ``mini gap'' structure is present.
As $D_F$ is increased
to $D_F=50$, a much
smaller gap remains, and multiple ripples rise to two
larger bound state peaks.
If $D_F$ is doubled again to $D_F=100$, the gap
disappears.
Upon increasing $D_F$ further to $D_F=200$, we see another
peak emerge and form the initial stages of the sawtooth-like
profile seen earlier in bulk systems (see Fig.~\ref{dos6}).
Thus we find that there exists a maximum thickness
for the normal metal $D_F\approx\Xi_0$, such that, if exceeded,
the gap in the normal side DOS disappears. The observed
filling in of the states originates from quasiparticles
with relatively large 
momenta parallel to the interface ($k_\perp\approx k_{FS}$).

In the right column of Fig.~\ref{dosnvary}, the local DOS 
in the superconductor at a distance
$Z=\Xi_0$ from the interface is shown. 
Here we do not spatially average
the local DOS, since we are in the bulk regime
and the DOS varies smoothly.
We  present the local
DOS at $Z=\Xi_0$, while
all 
other parameters take the values
used previously in the left column.
The top curve ($D_F=20$)
shows a widening of the gap, while the main peaks still 
remain below $\epsilon/\Delta_0=1$. The panel below
demonstrates the bound state peaks being pushed further
towards the Fermi level. For  $D_F=100$,
the single pair of peaks has moved inward even further as
a marked through develops at $\epsilon/\Delta_0 \approx 1$.
The effect is more pronounced in the bottom curve, where $D_F=4\Xi_0$,
and the de Gennes St.~James states have become smaller
than the main peaks that have formed near the gap edge, which eventually
develop into BCS-like peaks deeper within the superconductor. 

\begin{figure}[t]
{\epsfig{figure=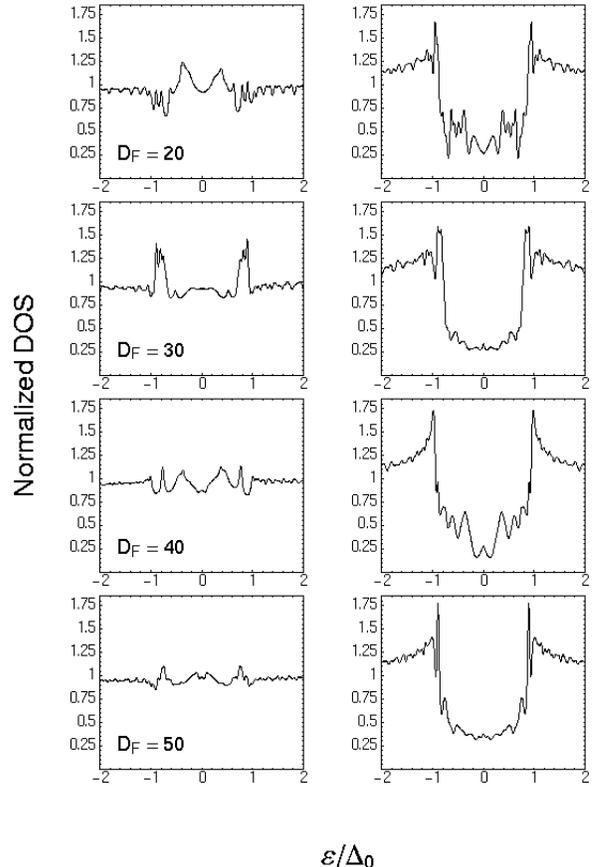,width=.45\textwidth}}
\caption{Local DOS as in the previous Figure,
but for $I=1/2$. From top to bottom each pair
of panels corresponds to $D_F=20,30,40,50$.
}  
\label{dosfvary}  
\end{figure}

We consider next the DOS when the non superconductor layer in 
the system is a ferromagnet. We  present results
at $t=0.02$, and $I=1/2$, as in the bottom
panel of Fig.~\ref{nvary}. In the magnetic
side, we spatially average the DOS
over its width $D_F$ as described above.
The result is shown in 
the left column of Fig.~\ref{dosfvary}. The top curve
exhibits two slightly asymmetric peaks at $\epsilon/\Delta_0\approx 0.2$.
The structure seen there
is washed out at larger $D_F$.
There is no gap in the ferromagnet DOS
shown but we found\cite{thesis} a
mini-gap when $D_F$ is small ($D_F\approx k_{FS}\xi_2$).
The local DOS at the point $Z=\Xi_0$ inside the superconductor
is illustrated  in the right column of Fig.~\ref{dosfvary}.
In the top curve ($D_F=20$), 
it is seen that the highermost
peaks are shifted slightly towards lower energies
($\epsilon/\Delta_0 \approx 0.9$)
compared with the bulk BCS result.
At even lower energies 
there is a relatively high number
of subgap states.
Upon increasing the ferromagnet's width to $D_F=30$, the coarse
structure seen previously becomes somewhat
smoothed out.
On the curve below ($D_F=40$),
numerous peaks have returned within the gap,
and then diminish again in the
the last curve
for  $D_F=50$. 
We find therefore that
the presence
of the magnet next to the superconductor 
results in  more prominent features in the DOS,
at  smaller $D_F$ values.

\begin{figure}[t]
{\epsfig{figure=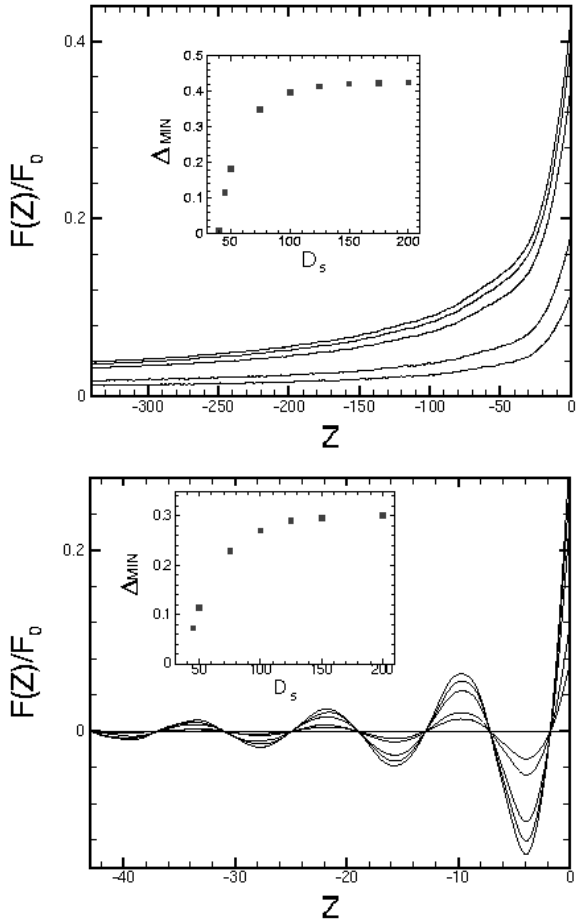,width=.45\textwidth}}
\caption{Pair amplitude for a structure consisting of a superconductor
of finite thickness $D_S$ adjoining a thick non-superconductor.
The main plot in the top panel shows the decay of the pair amplitude
in the normal metal ($I=0$) for values of $D_S=200,100,75,50,45$. The bottom
panel shows the pair amplitude at $I=1/2$ and the same geometry. The
values of $D_S$  are the same as in the top panel, and the amplitude
of the oscillations decays with decreasing $D_S$. The insets illustrate
the behavior of $\Delta_{\rm MIN}$ vs $D_S$ in each case.}  
\label{svary}   
\end{figure}

We now reverse the role of the two materials in the bilayer,
that is, we consider a very thick ``bulk'' ferromagnet  
(we take $D_F = 16 \Xi_0$ as was done above for $D_S$), in contact
with a finite superconductor layer. 
Temperature and other parameters are as in the
previous case. In order
to study fully the geometrical effects associated with
varying the superconductor thickness,
we shall consider a wide range of widths $D_S$,
taking $\Xi_0$ close to the  lower bound, since the superconductor ceases to maintain
pairing correlations when $D_S \lesssim \Xi_0$.

The top panel of
Fig.~\ref{svary} shows  the modification 
of the 
pair correlations in a bulk normal metal ($I=0$)
that occur as the width of the superconductor varies.
The top
curve ($D_S=4\,\Xi_0$) differs 
relatively little from the situation where both the normal
metal and superconductor were
in the bulk
(compare with Fig.~\ref{Fzero}). A decreasing trend is followed 
as $D_S$
decreases. 
The slow decay of $F(Z)$ 
away from the interface
is  adequately fit by Eqn.~(\ref{new}) for 
$D_S\ge1.5\,\Xi_0$, 
the only  modification being an overall $D_S$-dependent 
factor that reduces the amplitude.
The bottom two curves,
corresponding to $D_S=0.9\,\Xi_0, \Xi_0$ have an even slower
decay.
The inset depicts the corresponding change
in the pair potential at the interface,
$\Delta_{\rm MIN}$, as a function of $D_S$.
This inset emphasizes the fast rise
in the pairing correlations at the interface
when $D_S$ is on the scale of $\Xi_0$, and
it includes additional values of $D_S$ not 
presented in the main figure.

In Fig.~\ref{svary} (bottom panel) we show the damped
oscillations of $F(Z)$ within the ferromagnet 
($I=1/2$) for the same values of $D_S$ as in the top panel. 
The main effect
of changing $D_S$ is to reduce the 
amplitude of the oscillations while
their period remains, as expected, the same.
Their amplitude however, drops very markedly
when $D_S$ 
approaches $\Xi_0$.
This is illustrated in the inset,
where we display $\Delta_{\rm MIN}$ versus 
$D_S$. The essential behavior
is similar to that in the $I=0$ case in the other panel, whereby
$\Delta_{\rm MIN}$ changes the most for  $D_S<1.5\,\Xi_0$.
The overall magnitude is reduced, however, by the finite
value of the exchange energy.

\begin{figure}[t]
{\epsfig{figure=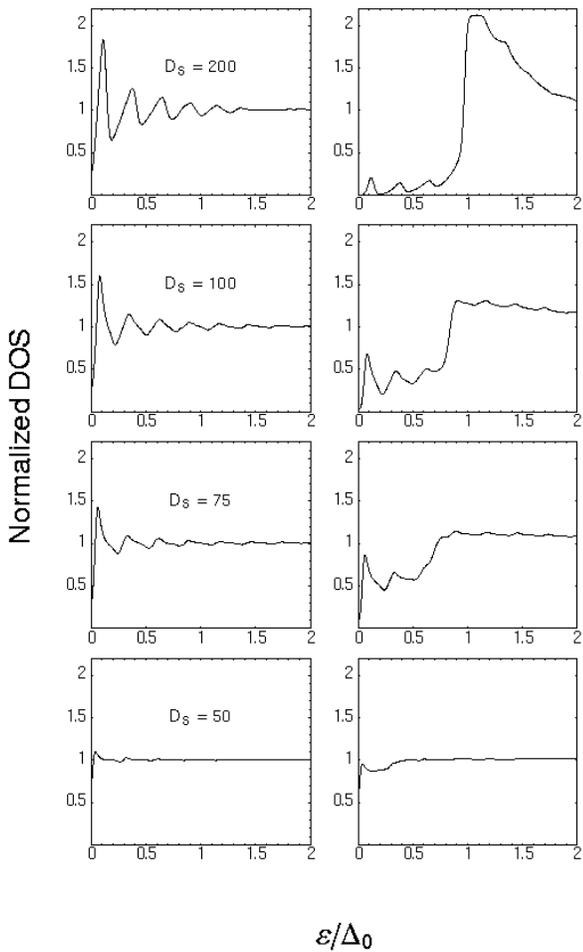,width=.45\textwidth}}
\caption{Left column: localized DOS at $Z=-100$
for the structure considered in
Fig.~\ref{svary}, at $I=0$ and superconductor widths
as indicated.
Right column: local DOS for the same structure,
averaged over one $\Xi_0$ 
(see text) from the end of the superconductor, 
for the same $D_S$ as in the left column.}  
\label{dossvary}    
\end{figure}

The local DOS is also sensitive to  the
spatial extent of the superconductor. 
The left column in Fig.~\ref{dossvary}
shows the normalized local DOS at  $Z=-100$ in the
$I=0$ limit, for several values of $D_S$.
Spatial averaging in this case is unnecessary.
For $D_S=200$ (top curve), it is evident
that this width is sufficient
for Andreev reflection to become
well established, and hence for the complete formation
of the structure  seen in the bulk case (see Fig.~\ref{dos6}).
At $D_S=2\Xi_0$, even though $D_S$ is  still larger
than $\Xi_0$, the DOS profile has changed its shape so that
the peaks slant in the opposite direction. The magnitude
of the  peaks has decreased slightly overall, but their
characteristic energy spacing $E_c$ remains the same with
the exception that the largest peak  has slightly shifted
towards smaller energies.
The next curve ($D_S=1.5\Xi_0$) shows how this trend extends further.
Finally, when $D_S=\Xi_0$, only a slight hint of structure remains.
These results reflect that 
the de Gennes St. James peaks arise mainly from Andreev reflection
at the normal metal superconductor interface, 
so that
when $D_S$ decreases, so does the minimum gap (see Fig.~\ref{svary} inset
in the top panel). 
Next, 
we examine the DOS in the superconducting
side (right column),  using the
same parameters as for the left side panels. 
We 
perform a spatial
average over one coherence length 
centered around $Z=D_S-\Xi_0/2$
(in analogy with the thin magnet case). Beginning with
the top curve and proceeding downwards, we see
a rapid filling in of subgap states.
As $D_S$ is decreased (lower curves), any remnant of subgap states 
becomes smeared
out due to the greater influence of the normal metal on
the superconductor for smaller $D_S$. 

\begin{figure}[t]
{\epsfig{figure=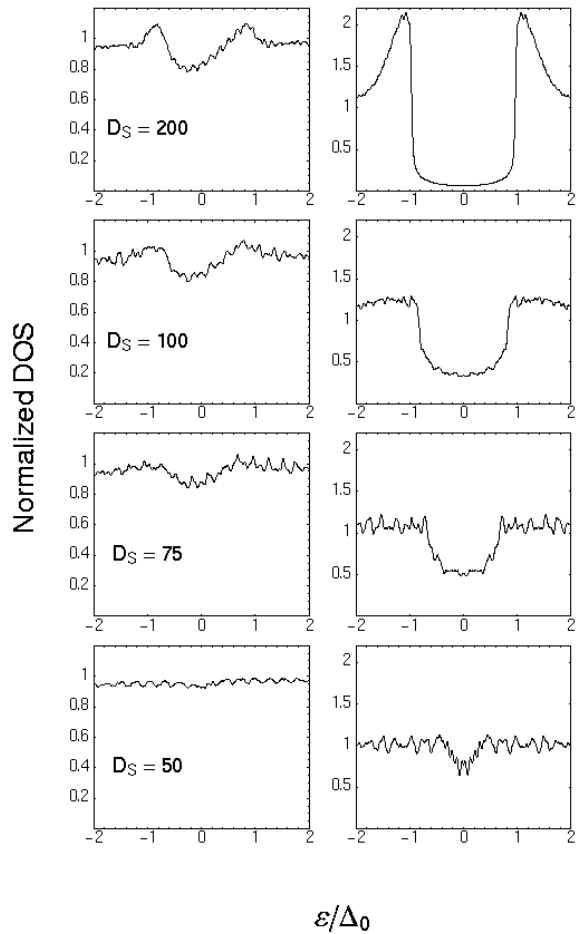,width=.45\textwidth}}
\caption{Local DOS for the structure
considered in Fig.~\ref{svary} with $I=1/2$.
Left column: localized DOS at $Z=-3$
for $I=1/2$ and  superconductor widths as labeled.
Right column: local DOS averaged over one $\Xi_0$ 
from the end of the superconductor for the same values 
of $D_S$ as in the corresponding left column panels.}  
\label{dos5svary}  
\end{figure}

We conclude this subsection by considering a finite exchange energy
of $I=1/2$ in the ferromagnet, with all
other material parameters being identical to
those in Fig.~\ref{dossvary}. 
As mentioned above, for
a bulk superconductor juxtaposed
to a bulk ferromagnet,
the oscillatory pair amplitude in the ferromagnet 
induces oscillations
in the DOS as a function of position within
the ferromagnet (see discussion of Fig.~\ref{dos1/21}).
Here, we wish to examine any modifications
on this behavior that may arise
from a  finite $D_S$.
To this purpose, we present in Fig.~\ref{dos5svary} (left column)
the local DOS at the position $Z=-3$ for several values
of $D_S$.
The top curve, corresponding to $D_S=4\Xi_0$, shows
a DOS profile with two rounded peaks near the gap edge,
while the minimum is at $\epsilon/\Delta_0=0$. This is fairly
similar  to what was seen at the same distance
in Fig.~\ref{dos1/21}.
To understand the behavior of the DOS for smaller $D_S$, we
recall the spatial dependence of the pair correlations in Fig.~\ref{svary}.
There it was found that the oscillations in $F(Z)$ did not undergo
a change in period as the superconductor width decreased. Rather, there was
a smooth reduction in  amplitude as $D_S$ decreased.
This suggests that  changes in the DOS with decreasing $D_S$
would behave similarly. The other curves in 
the left column of Fig.~\ref{dos5svary}
agree with this reasoning: the effect of reducing $D_S$ is to
lower  peaks and raise  minima, so that the leakage
of superconducting order is effectively eliminated
when $D_S=\Xi_0$. 

The  panels in
the right column of Fig.~\ref{dos5svary} show the corresponding DOS
in the superconductor, averaged over a distance
of one $\Xi_0$ in the usual way.
The top panel ($D_S=4\Xi_0$, so that the average 
is taken centered at
$Z=175$)
shows a clear but broad peak at $\epsilon/\Delta_0\approx 1$,
spreading  over
a significant energy range. 
This peak decays 
rapidly
while shifting to  smaller energies as $D_S$ decreases.
As  $D_S$ is reduced to $D_S=2\Xi_0$,
the states with energies $\epsilon/\Delta_0\lesssim 1$
fill in rapidly while the primary peak is flattened. This
trend continues until, at $D_S=\Xi_0$,  there
is hardly any evidence of the previous superconducting structure,
indicating  that superconductivity is nearly destroyed as the thickness
is down to one correlation length.

\subsection{Comparison  with experiment}
\label{compare}

We have seen above that the calculated self-consistent results for 
the physical
quantities depend
in a systematic way on a  number
of parameters, some of which are related to the materials
employed, while others are experimentally
adjustable.  While testing these systematic dependences is the
task of future experimental work, we will, in  this subsection,
compare already existing experimental data with
our theory.
We will use
data from direct local DOS measurements\cite{courtois,sillanpaa}
rather than results for indirectly
derived quantities.
Although we emphasize in this work the case
where the non-superconductor is a ferromagnet,
we will examine 
also the proximity effect 
when the normal metal is a ferromagnet.
We will thus compare our calculations
with the experimental data of Ref.~\onlinecite{courtois}
where DOS measurements were made
from the normal metal side, and with data
from Ref.~\onlinecite{sillanpaa}, 
where local DOS measurements
in the superconductor side of a magnet-superconductor structure
were taken. In this way,
we test our theory against spectroscopy data
obtained by probing
either side of the interface,
in the two  cases where the superconductor
is in proximity to either a normal metal or a
strong magnet. In making our  comparisons,
it is important to make pertinent choices for
the applicable input parameters,
as will  be discussed below.

Consider first the STM data of Ref.~\onlinecite{courtois},
where the superconductor (Nb) is in contact with a non-magnetic metal (Au).
The experimental configuration consisted of a thin layer of Au of thickness
$200$\AA, that capped off a  superconducting  Nb dot.
The Nb
had a smooth relief
resulting in
a thickness ($0-400$\AA) that decreased
away from its center. We model this
structure, as done in the experimental analysis\cite{courtois},
as a bilayer system comprised of a 
normal metal of
constant
width on a superconductor
of varying thickness 
in the $z$-direction,
in a manner similar to Sec.~\ref{structures}.
We take $k_{FM} = 1 \rm{\AA}^{-1}$, and we 
assume the normal metal to have
a width of 200 \AA.
The transition temperature
is $T_c=9$ K.
We assume the temperature to be 
$T=270$ mK,
a value which is slightly higher than the
experimental value $T=60$ mK. This is intended,
in the usual way,
to account for additional smearing
effects associated with the finite energy
resolution of the apparatus.
The interfacial barrier strength parameter ($H_B$) 
is taken to be
zero, which is
appropriate for the clean, highly transparent 
interface used, and 
the Fermi wavevectors
in the two materials are assumed to be equal.
Other parameter values used in our calculations
are the bulk Nb gap value $\Delta_0=2$ meV, which
is close to the experimental value\cite{courtois},
and the Debye cutoff parameter $\omega=0.03$,
the value of this parameter having little
effect on the results.
The modeling of the superconductor width
is less trivial since the Nb
dot in fact varies in size not only in the $z$ direction,
but
also in the transverse direction.
With this in mind,
we assume a superconductor thickness
varying from 50 to 150 Angstroms.
Because of possible nonuniformities in the
composition of the Nb dot, 
this quantity  should be viewed 
as an effective thickness that accounts
for any geometrical discrepancies between
our model and the experimental configuration,
and which  may  be interpreted to some extent as
a fitting parameter.
The final physical parameter needed
is the  coherence length $\xi_0$.
This parameter must be
identified here as an effective correlation
length to absorb
the inherent effects of
disorder in the system.
Measurements were taken at several points: some,
(which were labeled as points $a-d$ in Figure 1a
of Ref.~\onlinecite{courtois}) were on the
flatter part near the center of
the dot. The others, labeled $e-j$ in that Figure,
were in the sloping part near the edge.
Disorder effects are likely to be more prevalent
in the region in the latter points,
where  oxidation of the sample
surface has a more pronounced effect
on the superconducting order.
Therefore we have set at $\xi_0=100 \rm{\AA}$ 
in the region corresponding to
the points $e-j$,
while for in the region between
the points $a-d$ 
we take $\xi_0 = 200$\AA.
These two sets of points were recognized
as behaving differently in the
original experimental analysis\cite{courtois}.

\begin{figure}[ht!]
{\epsfig{figure=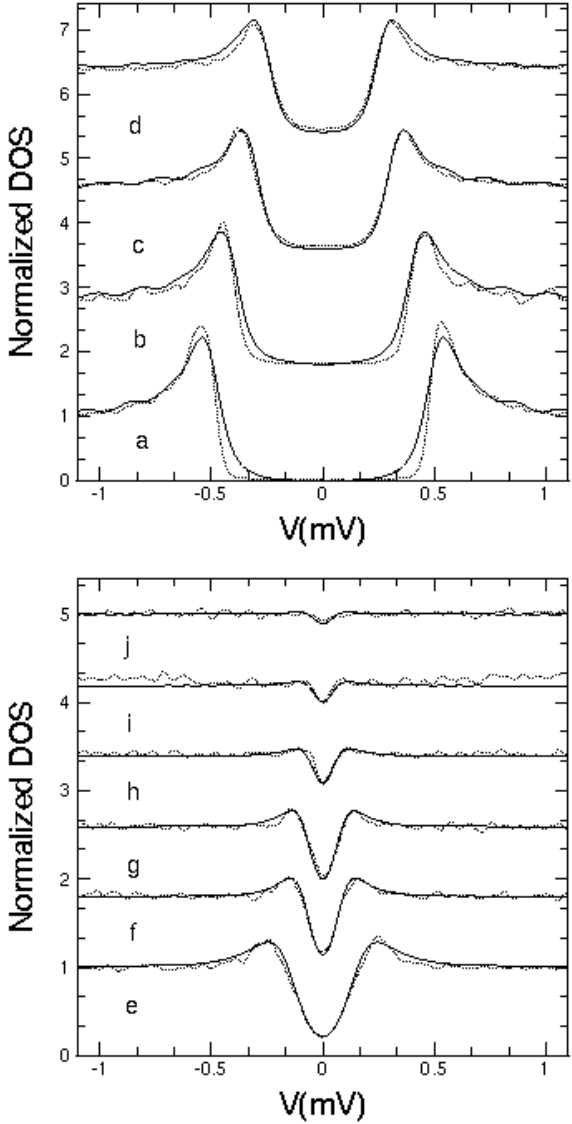,width=.45\textwidth}}
\caption{Comparison of  STM data from
Ref.~\protect\onlinecite{courtois} with theory. Solid curves:
theoretical results. Dotted curves:
data. Vertical
scales  shifted for clarity.
The labels $a$ through $j$ correspond to different
probe positions. See text for details.}
\label{courtdos} 
\end{figure}

The geometry studied is not quite that in our earlier results 
of Sec.~\ref{structures}, where
we varied the width of the superconductor in
contact with an {\it infinite} normal metal. Here, both
the normal metal and superconductor are effectively thin,
and additionally, the assumed value of the coherence length
is larger. Therefore, separate computations were required.
We present in Fig.~\ref{courtdos} the comparison of our 
results
(solid line) to the the experimental data.\cite{courtois}
The DOS is scaled to its normal metal limit,
with the curves  shifted by a constant for illustrative purposes. 
The energy is in the same voltage units as in the experiment.
Inspecting the spectra corresponding to the
points $a-d$ (labeled in our Figure as they were in the
experimental work),
Fig.~\ref{courtdos} demonstrates the excellent agreement between
our results and the data. 
In the top panel, a BCS-like gap is most evident for the 
position the location labeled $a$, and
as the effective superconductor width is decreased
from 150 \AA in  $a$ to 120 in $d$,
the gap becomes smaller, while the BCS-peaks
decrease in magnitude.
The location of  the peaks in the fits 
and  in the experimental data are seen to  essentially
agree.
A similar trend is seen for the remaining
probe locations in the bottom panel of Fig.~\ref{courtdos}, where
the effective coherence length and $D_S$
are smaller. The peaks move inwards while the previously 
empty
gap starts to fill in, with
an approximately linear rise near the Fermi level. This DOS
profile is 
observably different from that in the panel above, where the subgap DOS
had a U-shape compared to the V-shape here.  
It is remarkable that the level of agreement 
between theory and experiment is so high,
in that
the location of the peaks
as well as the origin of the
minima in the DOS match well over the entire
spatial range.
Thus we find that modeling
this particular experimental structure
as a bilayer is successful over
the entire spatial range. In Ref.~\onlinecite{courtois}
where it was found that a fit to all the data using
the Usadel equations was not possible
and that very different  physical  assumptions
had to be used for the U- and V-shaped portions.
This is unnecessary within the exact theory.

When the normal metal is a ferromagnet,
experimental studies on the proximity effect 
are more sparse.
The continuing advancement in nanofabrication techniques
however,
has made 
probing the electronic structure of
$F/S$
nanostructures experimentally accessible and some recent
good quality data is available.
We compare our theory with the tunnel
spectroscopy 
data of Ref.~\onlinecite{sillanpaa} obtained
through probing the local DOS in an Al
superconductor adjacent to a Ni ferromagnet.
Modifications
to the DOS in the superconductor are another
important aspect of the proximity effect
which provides useful information
regarding the influence of the
ferromagnet on superconducting correlations.
As predicted in Sec.~\ref{temp}
the local DOS
near the interface in the superconductor should be
substantially modified from the bulk BCS result.
It is then of great interest to see
how our results compare with the appropriate experiment.

To
test our theory against
the Al-Ni experimental data, we
must choose a  
set of
parameters appropriate
for the given bilayer.
Nickel itself is not a simple Stoner magnet with
parabolic bands. A nearly free electron
monovalent metal having the same saturation magnetization\cite{kittel} as Ni,
is easily seen to be to have a 
value of $I$ of about 0.5 and this is the value we will use.
We assume,
a transparent interface 
($H_B=0$) in accordance with 
the clean interface in the
experiment.\cite{sillanpaa}
We also take
$T_c=1.2$ K, $k_{FM}=0.5 {\rm \AA}^{-1}$, and in order 
to limit as much as possible the number of input parameters,
we keep the Fermi wave vectors the same, $\Lambda=1$.
For thick superconducting layers 
the relevant length which
governs the depletion of superconducting correlations
near the interface\cite{proximity} scales with $\xi_0$ whenever
$\xi_0 \gg 1$.
In the experimental work, distances were already
given in units of the correlation length, and this
makes it particularly convenient to compare
with theory.  
Experimental data is given at two distances:
one far from the interface and the other near to
it. The  precise distance from the tunneling probe
to the interface in the second case was somewhat uncertain, however, mainly
because of  the finite width of the probe. We 
take this position to be $2 \,\xi_0$,
which is similar 
to the value
estimated in the experimental analysis.\cite{sillanpaa}

\begin{figure}[ht!]
{\epsfig{figure=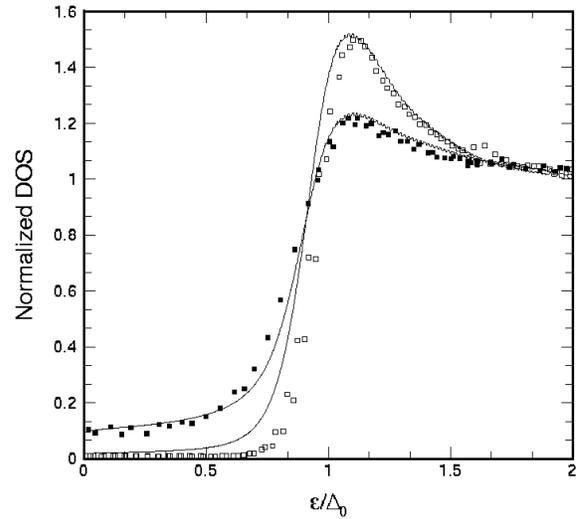,width=.45\textwidth}}
\caption{Comparison of experimental DOS data for Ni-Al
structure~\protect\cite{sillanpaa} with theory. The symbols represent 
data taken far from the interface (open symbols)
and near it (closed symbols). The curves are the theoretical
fits obtained as explained in the text.}
\label{sildos}
\end{figure}

Figure~\ref{sildos} shows a fit of our results (solid lines)
to the experimental data of Ref.~\onlinecite{sillanpaa}.
The vertical axis is the DOS scaled to
the normal state value, while the energies are
in dimensionless units of $\epsilon/\Delta_0$.
We account for
the influence  which single-charging effects have on the
DOS by
convolving
the  DOS calculated from ({\ref{tdos}) with the acceptance function $P(E)$ 
(as described in Ref.~\onlinecite{sillanpaa})
that gives the probability for the junction to absorb
an energy $E$.\cite{devoret}
The curve with the more prominent peak
corresponds 
to the DOS in the bulk, away from the influence of the
magnet, while the other curve is the DOS two coherence lengths from
the interface.
The open and closed symbols are the experimental
results, the former ones being for the bulk DOS. The  procedure 
employed to obtain these fits was the following:
first we determined the effective
temperature (to account for instrument resolution)
by fitting our results
to the experimental DOS near the interface, 
(closed symbols, solid curve). 
This resulted in 
a slightly raised effective temperature 
$T=980$ mK
(in contrast with the experimental value
of  $T=100$ mK).
Then, without any further changes,
we calculated the DOS at a distance of
$4 \,\xi_0$ from the interface (dashed curve).
This position should represent well the bulk
characteristics in the DOS, since as we
have seen previously,
at that point,
the influence
of the ferromanget on the superconducting
DOS is minimal.
No additional parameters were used to
obtain this second fit. The results are
clearly excellent: they have the correct
peak positions 
and relative magnitudes.
We therefore find again,
good overall
agreement with experiment in this more difficult case.

\section{conclusions}
\label{conclusions}

We have in this work  presented extensive results
for the pair amplitude and local DOS in heterostructures
involving superconductors and magnetic materials. These
results were obtained from numerical, self-consistent
solution of  the Bogoliubov-deGennes equations,
without approximations. We also discussed
the 
length scales characterizing the influence of the superconductor
on the ferromagnet and vice versa.

For heterostructures with geometric
dimensions larger than the relevant intrinsic
lengths, we have shown in detail how variation
of parameters such as temperature $T$,
Fermi wave vector mismatch $\Lambda$, interfacial
and scattering strength $H_B$,
(see Table~\ref{table1})
affected the pair amplitude and local DOS for
a wide range of exchange energies $I$.
For $I=0$,
and low $T$, we find that the pair amplitude 
in the normal metal
decays
approximately
as the inverse of the distance from the interface,
with the overall prefactor 
depending on $\Lambda$ and $H_B$. 
At higher temperatures,
and for $\Lambda=1$, $H_B=0$,
the decay markedly increases, 
and is
set by a smaller length scale $\xi_N(T)$. The exact
spatial decay of $F({\bf r})$ was found to be more complicated than 
a single
exponential. On the superconductor side,
we tested our results
for temperatures near $T_c$ and found 
agreement
with
standard Ginzburg Landau theory. We
also extracted the characteristic length of depletion $\xi_S(T)$
for intermediate and lower temperatures, 
something not previously done in a systematic way.
The  length scale (at low temperatures)
characterizing the 
decline of $F({\bf r})$ near the interface
was found to decrease with smaller $\Lambda$ or larger $H_B$.
The local DOS correlates with these results 
and displays a functional dependence on
$T$,
$\Lambda$, and $H_B$ as well.
We also systematically investigated the geometrical 
effects associated with
the finite size of either the
normal metal or superconductor and 
discovered that a small gap develops in the normal
metal DOS when its width is small. This  ``mini-gap'' decreases to zero when
the metal width is close to the coherence length, i.e., 
$D_F\approx\Xi_0$.

At finite values of $I$, 
there are two characteristic proximity length
scales
$\xi_2$ and $\xi_1$ in the ferromagnet, 
governing respectively the spatial period
of the damped oscillations in
$F({\bf r})$, and the sharp decay at
the interface. Both lengths vary approximately inversely with $I$, 
independent of temperature.
The amplitude of the damped oscillations 
however, decreases
with increasing temperature.
Mismatch of the Fermi energies
increases the decay of the oscillations.
A finite barrier strength
has no effect on the period either,
but does reduce the amplitude by a factor 
that scales linearly with $H_B$.

The damped oscillatory behavior
in the ferromagnet induces a corresponding 
spatial modulation in the local DOS.
For a transparent interface,
the  DOS in the superconductor side (for $I>1/4$),
exhibited a reduction in
the usual BCS peaks, with a finite number of
states within the gap, the number depending
on the exchange field and location within the superconductor.
For small exchange fields of order of $I\simeq\Delta_0/E_{FS}$,
a significant subgap structure emerged and at exactly $I=\Delta_0/E_{FS}$,
a resonance phenomenon occurred in which the BCS
peaks became significantly enhanced.
We found there exists a long range spin splitting in the
superconductor, extending over several coherence lengths.
A nontrivial  behavior
of $\delta N$ was found as a function of $I$ and
the mismatch parameter $\Lambda$:
when the point $E_{F\uparrow}=E_{FS}$ is crossed,
the spin splitting
becomes very short ranged, as found
in Ref.\onlinecite{proximity}.

Finally, we have compared our results with two sets of experimental
data for the local DOS, corresponding to two different
values of $I$ and to measurements taken either from the superconductor
or the non-superconductor side of the heterostructure. In both cases
we found, using reasonable values for the material and geometric parameters,
very good agreement between theory and experiment.

The number of relevant parameters involved is so large,
and the variety of behaviors so rich, that even an extensive
study such as this one must concentrate on the
highlights and leave most of parameter space unexplored. It is clear
however that the machinery developed here can be readily applied to
most actual experimental situations. We
hope that this paper will stimulate future experimental work and facilitate
the analysis of the resulting data.

\acknowledgments
We thank H. Courtois and  M.A Sillanp\"{a}\"{a}  for providing us with their
data.

%
%

%
%
}
\end{twocolumn}

\begin{references}  
\bibitem[\dagger]{klaus} Electronic address: khalter@physics.umn.edu  
\bibitem[*]{oriol} Electronic address: otvalls@tc.umn.edu  
\bibitem{kraus} P.A. Kraus, A. Bhattacharya and A. M. Goldman, \prb {\bf 64},
220505 (2001).
\bibitem{grig} A. N. Grogorenko {\it et al.} \prb {\bf 63}, 052504 (2001).
\bibitem{sillanpaa} M.A. Sillanp\"{a}\"{a}, T.T. Heikkila, R.K. Lindell,
and P.J. Hakonen, Europhys Lett {\bf 56}, 590 (2001).
\bibitem{parks} G. Deutscher and P.G. de Gennes, in  
{\it Superconductivity}, edited by R.D. Parks 
(Marcel Dekker, New York, 1969), p. 1005. 
\bibitem{falk}  D.S. Falk, Phys. Rev.  
{\bf 132}, 1576 (1963).
\bibitem{bdg} P.G. de Gennes, {\it Superconductivity of Metals 
and Alloys} (Addison-Wesley, Reading, MA, 1989).
\bibitem{branko2} B.P. Stojkovi\'c, and O.T. Valls, 
\prb {\bf 50}, 3374 (1994).
\bibitem{hara} J. Hara, M. Ashida, and K. Nagai, \prb
{\bf 47}, 11263 (1993).
\bibitem{demler}  E.A. Demler, G.B. Arnold, and M.R. Beasley, \prb  
{\bf 55},15 174 (1997).
\bibitem{blum} Y. Blum, A. Tsukernik, M. Karpovski, A. Palevski
cond-mat/0203408.
\bibitem{prokic} V. Prokic, A.I. Buzdin, and
L.Dobrosavljevic-Grujic, \prb {bf 59} 587 (1999).
\bibitem{bergeret} F.S. Bergeret, A.F. Volkov, and K.B. Efetov,
cond-mat/0106510.
\bibitem{bulaevskii} L.N. Bulaevskii, V. V. Kuzii 
JETP Lett. {\bf 25}, 290 (1977).
\bibitem{buzdin82} A.I. Buzdin, L.N. Bulaevskii, S.V.
Panyukov, JETP Lett. {\bf 35}, 178 (1982).
\bibitem{buzdin92} A.I. Buzdin, B. Bujicic, 
M. Yu Kupriyanov, JETP Lett. {\bf 74}, 124 (1992).
\bibitem{fulde} P. Fulde and A. Ferrell, Phys. Rev.  {\bf 135}, A550 (1964).
\bibitem{larkin} A. Larkin and Y. Ovchinnikov, 
Sov. Phys. JETP {\bf 20}, 762 (1965). 
\bibitem{radovic}  Z. Radovic, {\it et al.}  \prb {\bf 44},
759 (1991).
\bibitem{tagirov} L.R. Tagirov, Physica C {\bf 307}, 145 (1998).
\bibitem{fominov2} Y.V. Fominov, N.M. Chtchelkatchev, and
A.A. Golubov, cond-mat/0202280.
\bibitem{kontos2} T. Kontos {\it et al.} cond-mat/0201104 (2002).
\bibitem{ryazanov} V.V. Ryazanov {it et al} \prl {bf 86}, 2427
(2001).
\bibitem{gyorffy} M. Krawiec, B.L. Gyorffy, and J.F. Annett,
cond-mat/0203184.
\bibitem{fominov} Ya. V. Fominov and M.V. Feigelman,
\prb {\bf 63}, 094518 (2001).
\bibitem{baladie} I. Baladie and A. Buzdin, \prb
{\bf 64} 224514.
\bibitem{zareyan} M. Zareyan, W. Belzig, and Yu. V. Nazarov,
cond-mat/0107252.
\bibitem{usadel} K.D. Usadel, \prl {\bf 25}, 507 (1970).
\bibitem{eilen} G. Eilenberger, Z. Phys. {\bf 214}, 195 (1968).
\bibitem{tanaka}  Y. Tanaka and M. Tsukada, \prb  
{\bf 42},2066 (1990).
\bibitem{zv} I. \v{Z}uti\'c and O.T. Valls, \prb {\bf 61}, 1555 (2000).
\bibitem{vodo} B.P. Vodopyanov, {it et al} Physica C, {\bf 366},31
(2001).
\bibitem{bourgeois} O. Bourgeois, {\it et al}, \prb {\bf 63}, 
064517 (2001).
\bibitem{aarts} J. Aarts, {\it et al}, \prb {\bf 56}, 
2779 (1997).
\bibitem{proximity} K. Halterman and O.T. Valls,
\prb {\bf 65}, 014509 (2002).
\bibitem{ting} J-X. Zhu and C.S. Ting, \prb {\bf 61}, 1456, (2000).
\bibitem{vecino} E. Vecino, A. Martin-Rodero and A. Levy Yeyati,
\prb {\bf 64}, 184502 (2001).
\bibitem{courtois} N. Moussy, H. Courtois, and B. Pannetier,
Europhys. Lett. (2001).
\bibitem{ketterson} J.B. Ketterson and S.N. Song, {\it Superconductivity},
Cambridge University Press, (1999). See Ch. 41.
\bibitem{pv} S.W. Pierson and O.T. Valls , Phys. Rev. {\bf B45}, 2458 (1992).
\bibitem{stjames}  P.G. de Gennes and D. St.-James, Phys. Lett.  
{\bf 4}, 151 (1963). 
\bibitem{bruder} S. Pilgram, W. Belzig, and C. Bruder,
\prb {\bf 62}, 12462 (2000).
\bibitem{thesis} K. Halterman, Ph.D. thesis, University of Minnesota,
(2002).
\bibitem{fazio} R. Fazio, and C. Lucheroni, Europhys. Lett. {\bf 45}, 707 (1999).
\bibitem{btk} G.E. Blonder, M. Tinkham, and T.M. Klapwijk, \prb
{\bf 25}, 4415 (1982).
\bibitem{sv} B.P. Stojkovi\'c, and O.T. Valls, \prb {\bf 47}, 5922 (1993). 
\bibitem{kittel} See Chapter 15 in C. Kittel {\it Introduction to Solid
State Physics}, 7th edition, Wiley, New York (1996).
\bibitem{gueron} S. Gueron, {\it et al}, \prl {\bf 77}, 3025 (1996).
\bibitem{devoret} G.-L. Ingold and Yu. Nazarov, in {\it Single Charge Tunneling},
ed. by M.H. Devoret and H. Grabert, Plenum, N.Y. (1992).
\bibitem{belzig}  W. Belzig, C. Bruder, and G. Schon, \prb  
{\bf 54}, 9443 (1996).
\bibitem{golu} A.A. Golubov, Physica C, {\bf 326}, 46 (1999).

\end{references}
\end{document}